\documentclass[prd,reprint,
nofootinbib,notitlepage,aps,tightenlines,
preprintnumbers,amsmath,amssymb,amsfonts,showpacs,superscriptaddress]{revtex4-2}

\usepackage{caption,subcaption}
\usepackage{mathrsfs} 
\usepackage{booktabs}
\usepackage[pdfencoding=auto,hyperindex,breaklinks]{hyperref}
\usepackage{graphicx,epstopdf}
\usepackage{mathtools}
\usepackage{enumitem}
\usepackage{caption,subcaption}
\usepackage{natbib,bm}
\usepackage[usenames,dvipsnames]{color}
\usepackage{slashed}   
\usepackage{comment}
\usepackage{multirow}
\usepackage{ulem} 
\usepackage{chngcntr}

\def\be{\begin{equation}}
\def\ee{\end{equation}}
\def\ba{\begin{eqnarray}}
\def\ea{\end{eqnarray}}
\def\ge{\mathrel{\raise.3ex\hbox{$>$\kern-.75em\lower1ex\hbox{$\sim$}}}}
\def\la{\mathrel{\raise.3ex\hbox{$<$\kern-.75em\lower1ex\hbox{$\sim$}}}}

\def\thesection{\arabic{section}}
\def\theequation{\arabic{equation}}
\def\simgt{\mathrel{\raise.3ex\hbox{$>$\kern-.75em\lower1ex\hbox{$\sim$}}}}
\def\simlt{\mathrel{\raise.3ex\hbox{$<$\kern-.75em\lower1ex\hbox{$\sim$}}}}

\newcommand{\nc}{\newcommand}

\nc{\gone}{\bar g_{\pi NN}^{(1)}}
\nc{\gzero}{\bar g_{\pi NN}^{(0)}}
\nc{\al}{\alpha}
\nc{\ga}{\gamma}
\nc{\de}{\delta}
\nc{\ep}{\epsilon}
\nc{\ze}{\zeta}
\nc{\et}{\eta}
\nc{\ka}{\kappa}
\nc{\rh}{\rho}
\nc{\si}{\sigma}
\nc{\ta}{\tau}
\nc{\up}{\upsilon}
\nc{\ph}{\phi}
\nc{\ch}{\chi}
\nc{\ps}{\psi}
\nc{\om}{\omega}
\nc{\Ga}{\Gamma}
\nc{\De}{\Delta}
\nc{\La}{\Lambda}
\nc{\Si}{\Sigma}
\nc{\Up}{\Upsilon}
\nc{\Ph}{\Phi}
\nc{\Ps}{\Psi}
\nc{\Om}{\Omega}
\nc{\ptl}{\partial}
\nc{\del}{\nabla}
\nc{\ov}{\overline}
\nc{\newcaption}[1]{\centerline{\parbox{15cm}{\caption{#1}}}}
\nc{\us}{U(1)$_S$}

\nc{\RN}{Reissner-Nordstr\"{o}m}
\nc{\Sch}{Schwarzschild}

\newcommand{\ehat}[1]{\hat e_{(#1)}}
\newcommand{\bl}[1]{\partial_{#1}}
\newcommand{\gmunu}[1]{g_{#1}}
\newcommand{\rthO}[1]{\biggr . #1 \biggr |_{(r_o, \theta_o)}}

\newcommand{\gM}[0]{\gamma_{\text{M}}}
\newcommand{\gK}[0]{\gamma_{\text{K}}}

\def\beq{\begin{equation}}
\def\eeq{\end{equation}}
\def\bmat{\begin{displaymath}}
\def\emat{\end{displaymath}}
\def\bear{\begin{eqnarray}}
\def\eear{\end{eqnarray}}
\def\ba{\begin{eqnarray}}
\def\ea{\end{eqnarray}}
\def\bery{\begin{array}}
\def\ery{\end{array}}
\def\bit{\begin{itemize}}
\def\eit{\end{itemize}}
\def\ben{\begin{enumerate}}
\def\een{\end{enumerate}}
\def\btab{\begin{tabular}}
\def\etab{\end{tabular}}
\def\btbl{\begin{table}}
\def\etbl{\end{table}}
\def\bfig{\begin{figure}[htb]}
\def\efig{\end{figure}}
\def\bpic{\begin{picture}}
\def\epic{\end{picture}}

\def\nnl{\nonumber \\}

\def\ga{\mathrel{\raise.3ex\hbox{$>$\kern-.75em\lower1ex\hbox{$\sim$}}}}
\def\la{\mathrel{\raise.3ex\hbox{$<$\kern-.75em\lower1ex\hbox{$\sim$}}}}
\def\gappeq{\mathrel{\rlap {\raise.5ex\hbox{$>$}}
{\lower.5ex\hbox{$\sim$}}}}
\def\lappeq{\mathrel{\rlap{\raise.5ex\hbox{$<$}}
{\lower.5ex\hbox{$\sim$}}}}

\def\gyr{{\rm \, G\kern-0.125em yr}}
\def\mev{{\rm \, Me\kern-0.125em V}}
\def\gev{{\rm \, Ge\kern-0.125em V}}
\def\tev{{\rm \, Te\kern-0.125em V}}

\begin{document}

\title{Null orbits and shadows in the Ernst-Wild geometry: insights for black holes \texorpdfstring{\\}{ } immersed in a magnetic field}

\author{Kate J. Taylor}
\affiliation{Department of Physics and Astronomy, University of Victoria,
Victoria, BC V8P 5C2, Canada}
\author{Adam Ritz}
\affiliation{Department of Physics and Astronomy, University of Victoria,
Victoria, BC V8P 5C2, Canada}

\date{\today}

\begin{abstract}
\noindent

We investigate the null geodesics, in particular the stable and unstable light rings and shadows, of a Kerr-Newman black hole immersed in an asymptotically uniform magnetic field as described by the Ernst-Wild (Melvin-Kerr-Newman) spacetime. 
Through numerical ray tracing, we demonstrate that both the black hole rotation and the magnetized Melvin geometry impact the light rings and shadows non-trivially and in compensating ways. In addition, we use a perturbative expansion in the magnetic field $B$ to analyze the deviation of the observable shadow relative to the Kerr result analytically, and determine connections between Lyapunov exponents for light ring instabilities and quasinormal modes in the eikonal limit. 
\end{abstract}
\maketitle
\section{Introduction}\label{sec:Introduction}
The ongoing LIGO-Virgo-KAGRA observations of gravitational waves from binary black hole (BH) mergers \cite{LIGOScientific:2016lio} and the direct observation of the supermassive BH candidates M87$^*$ and SgrA$^*$ by the Event Horizon Telescope (EHT) \cite{EventHorizonTelescope:2019dse,EventHorizonTelescope:2022xqj} have opened up a new observational window on black holes and nonlinear gravity. The prospect of space missions \cite{Johnson:2019ljv,Gralla:2020srx} and potential next-generation observatories (such as LISA, GRAVITY+, next-generation EHT, Cosmic Explorer, Black Hole Explorer,  Einstein Telescope) \cite{Buoninfante:2024oxl} providing higher precision probes of the near-horizon region, mass, spin, and photon ring structure of supermassive black holes demonstrates the importance of advancing the theoretical analysis of environmental effects impacting this regime. Such effects include accretion, local magnetospheres, and the plasma environment, not to mention potential sources of new physics \cite{Buoninfante:2024oxl}. Indeed, it has been emphasized \cite{Gralla:2020pra} that assuming the validity of general relativity and inferred black hole mass provides an ideal starting point to test higher-precision data for the impact of environmental effects. 

The magnetosphere is a primary quantity in the astrophysical black hole environment \cite{Wald:1974np,Blandford:1977ds},
playing a central role in realistic models of black energy production, such as the Blandford-Znajek process thought to power relativistic jets. A recent EHT analysis further corroborates the presence of strong magnetic fields near M87$^*$ \cite{EventHorizonTelescope:2021srq}. Magnetic fields naturally impact the trajectories of charged particles in the neighbourhood of the black hole, while the impact on neutral particles, and light in particular, is more subtle. However, given the expectation that magnetic fields may be present more generally in the environment of supermassive black holes, we are motivated to explore how the back-reaction of a magnetosphere on the black hole can impact null geodesics and the structure of photon rings and shadows. In particular, we aim to explore the interplay of the magnetic field with rotation in determining the shadow phenomenology. In this context, we will consider shadows both in the mathematical sense, bounded by the critical curve(s) describing the image of the photon shell, and also in the observable sense as the central darkness caused by the presence of a BH. These two notions do not precisely coincide \cite{Lupsasca:2024xhq}.

To analyze these questions, we require a model spacetime that incorporates the back-reaction of the magnetosphere on the black hole and in addition is physically consistent with the poloidal topology of fields induced for example by accretion. Exact Einstein-Maxwell solutions of this type that are asymptotically flat are not known, but the Ernst-Wild (EW) solution \cite{ErnstWild:1976Metric} incorporates the back-reaction of a poloidal magnetic field which is asymptotically uniform. The Ernst-Wild geometry, characterized by the magnetic field strength $B$, is obtained via a Harrison transformation on a seed Kerr-Newman black hole, and thus is not asymptotically flat, tending to a Melvin-like geometry \cite{Bonnor:1954,Melvin:1965zza} at large radius. Nonetheless, since known astrophysical magnetic fields are relatively weak in Planck units, the black hole and its near-horizon region can be parametrically isolated from the large-$r$ asymptotic Melvin-like domain. We will make use of this scale separation in the small-$B$ limit to analyze the weakly-magnetized, near-horizon Ernst–Wild geometry and compare it to the corresponding behavior expected from the default Kerr background.

In purely theoretical terms, relaxing the requirement of asymptotic flatness is perhaps the simplest mathematical way to bypass black hole no-hair theorems and generate new exact solutions to the Einstein-Maxwell equations. As a result, the structure of null geodesics can differ substantially from the expectations of the Kerr-Newman family, which exhibits only unstable light rings. Indeed, the Melvin geometry itself allows for stable light rings. Consequently, when a black hole is embedded in the Melvin spacetime, the observer sees the black hole shadow displayed panoramically along the equator for sufficiently large magnetic field strength. This is in stark contrast to asymptotically-flat Kerr black holes and illustrates a dramatic deviation from the astrophysically observed shadow. In addition, for sufficiently large magnetic fields, there are no light rings outside the horizon \cite{Junior_2021}. Of course, this is only true for parametrically large magnetic fields (compared to the black hole mass). In contrast, for weakly-magnetized systems, where the Melvin-like geometry is parametrically separated at large radius, we find instead a smooth deviation from the Kerr shadow as a function of $B$, allowing a quantitative analysis of the deviation.

In this work we will explore both the light ring structure and black hole shadows for magnetized geometries. We extend the ray-tracing analysis to the full Ernst-Wild black hole, (previously explored in \cite{Wang_2021} for the case with vanishing Kerr-Newman charge $q$ corresponding to a non-zero total electric charge $\mathcal{Q}$). We explore the impact of two particular values of the seed Kerr-Newman charge $q$: one of astrophysical motivation that ensures vanishing total electric charge $\mathcal Q=0$ and one of theoretical importance that removes non-trivial pathological issues present in the ergoregion \cite{Gibbons:2013EW}. We focus on the weakly-magnetized domain, which is of astrophysical interest, where the impact of the non-asymptotically flat nature of the Melvin-like geometry at large radius is negligible. We find an interesting interplay between the effects of black hole spin and magnetic field on the distortion of the Kerr shadow. We analyze these effects quantitatively by expanding perturbatively in $B$. By taking a further perturbative expansion in BH spin, we are also able to study quantitatively the instabilities of the unstable light ring as a function of $B$ by computing quasinormal modes in the eikonal limit.

The remainder of the paper is structured as follows. Section~\ref{sec:EWMetric} reviews the properties of the Ernst-Wild spacetime and its characteristic  geometry, including features of the apparent horizon. 
Section~\ref{sec:Light rings and equatorial geodesics} considers the effective potential for null geodesics and the light ring structure in the equatorial plane, considering the impact of three values of the seed Kerr-Newman charge $q$ and providing a comparison to the existing literature. 
Section~\ref{sec:RayTracing} introduces the backward ray-tracing technique and investigates the shadow cast by the Ernst-Wild black hole for a variety of parameters, including a quantitative analysis of the deviation of the shadow from Kerr for weak magnetization. Section~\ref{sec:EWinPerturbativeRegime} focuses on the slowly-rotating and weakly-magnetized Ernst-Wild black hole, and quantifies the impact of the magnetic field on the photon shell analytically within a perturbative framework. This section
determines characteristic properties of the unstable light ring - the orbital frequency and the Lyapunov exponent - as a perturbative expansion in $B$ and considers their connection to the quasinormal mode (QNM) spectrum determined by a WKB analysis in the eikonal limit. The QNM spectrum is validated by extending the WKB approach to higher order \cite{Iyer:1986np,Seidel:1989bp} and via comparison with the continued fraction approach described in \cite{Taylor:2024duw}. 

Throughout this work, we use geometric units $G=c=1$ and a Gaussian system with $4 \pi \epsilon_0 = \mu_0/4 \pi =1$. The metric signature is $(-,+,+,+)$ and Greek indices run from 0 to 3.

\section{General Ernst-Wild solution}\label{sec:EWMetric}
\begin{figure*}[t]
    \centering
    \includegraphics[width=\linewidth]{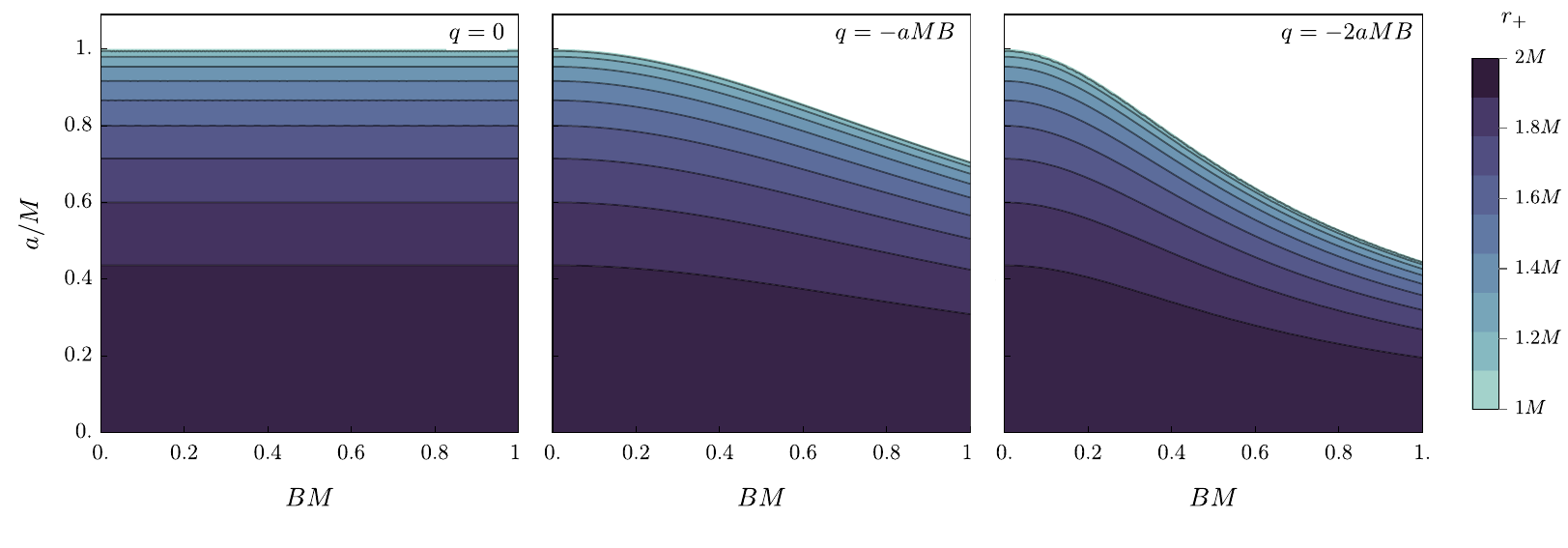}
    \caption{The event horizon for the Kerr-Newman seed metric for $q=(0, -a M B,-2a M B)$. The introduction of the non-zero charge parameter eliminates some of the parameter space as the values of $a$ and $B$ increase, as expected for a Kerr-Newman black hole as it saturates the extremal bound. }
    \label{fig:extremalLimit}
\end{figure*}
The Ernst-Wild solution arises from the use of a Harrison transformation \cite{Harrison:1968HarrisonTransformations} (see e.g. \cite{Gibbons:2013EW}) on a seed Kerr-Newman metric with parameters $(M,a,q,p)$, which determine the mass, angular momentum, electric  and magnetic charges of the geometry respectively. The line element can be written in the form,
\begin{equation}
    \label{e:MagnetizedKNMetric}
    \begin{split}
        	ds^2 &= \varrho ^2\, \Lambda \biggr [ -\frac{\Delta}{\Sigma}\,dt^2 + \left ( \frac{dr^2}{\Delta} + d \theta^2 \right ) \biggr ] \\ 
            &~~~~~~~~~~~+ \frac{\Sigma\, \sin^2 \theta}{\varrho ^2\, \Lambda} (\Lambda_0\, d\phi - \varpi dt )^2,\\
            A_\mu &= \Phi_0\, dt + \Phi_3 (\Lambda_0\, d \phi - \varpi dt ),
    \end{split}
\end{equation}
where
\begin{equation}
    \begin{split}
        \Delta &= r^2 - 2 M r +  a^2 + q^2+p^2, \\
        \varrho^2 &= r^2 +  a ^2  \cos^2 \theta, \\
	    \Sigma &= \left(r^2+a^2\right)^2-a^2 \Delta  \sin ^2 \theta.  \\
    \end{split}
\end{equation}
The functions $\Delta,\, \varrho^2$ and $\Sigma$ are identical to the standard Kerr-Newman black hole quantities. However, $\Lambda,\, \varpi,\, \Phi_0$ and $\Phi_3$ (given explicitly in Appendix~\ref{appendix:Full_EW}) are lengthy functions that arise from the magnetizing transformation procedure. We note that the parameter $\Lambda_0 = \left .\Lambda(r, \theta)\right |_{\theta=0}$ and is used to correct the conical singularity as shown by Hiscock in \cite{Hiscock:1981EWFixed} and is given by 
\begin{equation}
    \Lambda_0 =1+ a^2 B^4 M^2 +2 a B^3 M q+\frac{B^4 q^4}{16}+\frac{3 B^2 q^2}{2}.
\end{equation}

This fully back-reacted exact geometry has been analyzed in some detail in \cite{Gibbons:2013EW,Gibbons:2013dna,Astorino_2015,Astorino_2016,Brenner_2021,Booth_2015}. In what follows, we set the magnetic charge $p=0$ and work only with electric charge $q$. Note that the Kerr-Newman metric can be obtained by setting $B=0$, the Kerr metric can be obtained by setting $B=0$ and $q=0$, the Ernst metric can be obtained by setting $a=0$ and $q=0$ and the Schwarzschild metric follows by setting $B=0$, $a=0$ and $q=0$. 

The event horizons of the Ernst-Wild black hole are located at 
\begin{equation}
    r_\pm = M \pm \sqrt{M^2 -a^2 -q^2},
\end{equation}
and the radius of the ergosphere $r = r_e(\theta)$ is determined by 
\begin{align}
    g_{tt} = - \varrho ^2\, \Lambda\, \frac{\Delta}{\Sigma } + \frac{ \Sigma\,  \sin ^2\theta}{ \varrho ^2\, \Lambda }\, \varpi ^2 =0.
\end{align}
The ergoregion is enclosed by the ergosphere and the outer event horizon, that is $r_+ < r < r_e$. In the limit $B \rightarrow 0 $, the ergosphere radius reduces to the usual Kerr result $r_e = M - \sqrt{M^2 -a^2 \cos^2 \theta}$. 
\begin{figure}[b]
    \centering
    \includegraphics{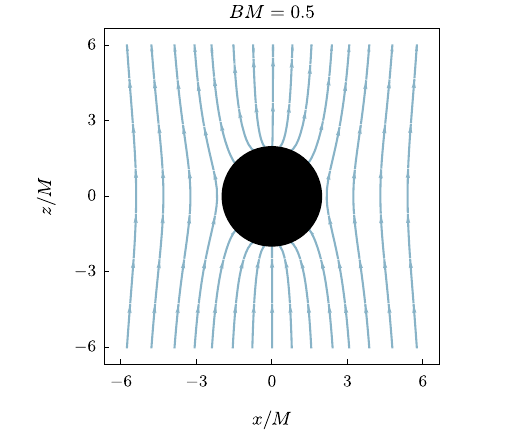}
    \caption{Magnetic field of a non-rotating Ernst-Wild black hole of mass $M$ with the magnetic field strength equal to $BM =0.5$. }
    \label{fig:ErnstBField}
\end{figure}

For illustration, we exhibit the poloidal form of the magnetic field most conveniently by setting $a=0$ in the electromagnetic potential given in Appendix~\ref{appendix:Full_EW}. The components of the magnetic field for the non-rotating Ernst black hole then reduce to
\begin{equation}
\begin{split}
    B_r &= \frac{B \cos\theta}{\left ( 1+ \frac{1}{4}B^2 r^2 \sin^2 \theta\right )^2},\\
    B_\theta &= \frac{B r \sin^2 \theta}{\left ( 1+ \frac{1}{4}B^2 r^2 \sin^2 \theta\right )^2},
\end{split}
\end{equation}
and are plotted in Fig.~\ref{fig:ErnstBField} for magnetic field strength $BM =0.5$. We note that for matter-free solutions of the Einstein-Maxwell equation, notably, the addition of rotation changes this picture by acting to exclude the magnetic flux from the horizon. This ``Meissner effect" fully excludes the magnetic field lines from the horizon in the extremal limit, as first observed in the perturbative Wald solution \cite{Wald:1974np}. 
\subsection{Conserved charges}
\begin{figure*}[t]
    \centering
    \includegraphics[width=0.9\linewidth]{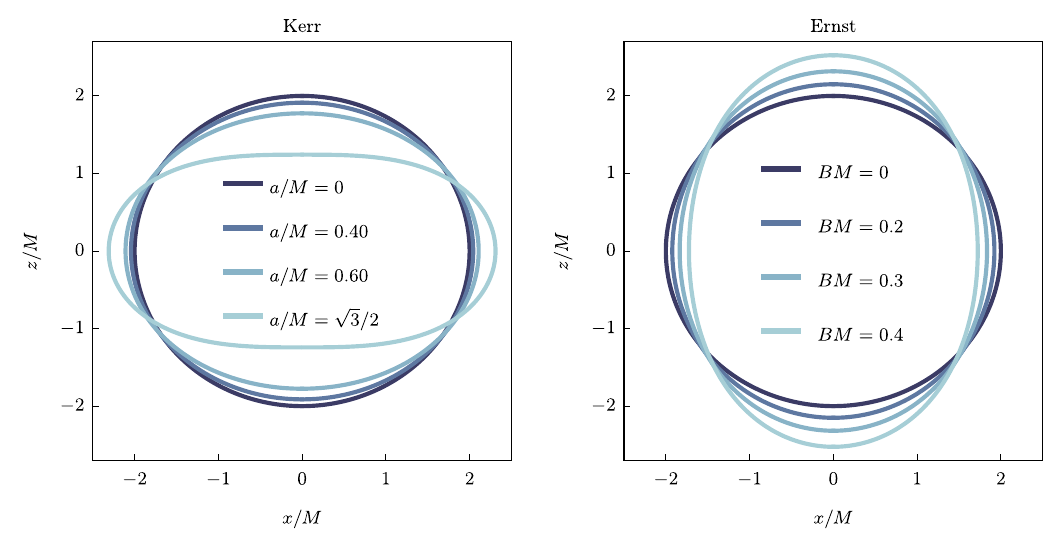}
    \caption{The apparent horizon geometry of the Kerr (left) and Ernst (right) black hole for four values of the spin $a/M$ and magnetic field $BM$, respectively. Both start out as the Schwarzschild geometry for $a/M = BM =0$ (darkest spherical curve). As the spin increases, the embedding surface becomes increasingly oblate, while as the magnetic field increases the embedding surface becomes increasingly prolate. }
    \label{fig:ernstKerrAH}
\end{figure*}
To understand the effects of the Harrison transformations, it is useful to write the observable electric charge $({\mathcal Q})$, angular momentum $({\mathcal J})$ of the Ernst-Wild solution in terms of the seed Kerr-Newman metric quantities $(q,\, a,\, M)$ \cite{Gibbons:2013EW,Gibbons:2013dna, Astorino_2015,Astorino_2016},
\begin{align}
    \label{e:QEW}
    \mathcal{Q} &=q + 2 a MB - \frac{1}{4}q^3 B^2, \\
    \mathcal{J} &= \ a M - q^3B -\frac{3}{2}q^2 a M B^2 - q \left (2a^2 M^2 +\frac{1}{4}q^4 \right )B^3 \nnl
    &\;\;\;\;- a M \left ( a^2 M^2 +\frac{3}{16}q^4\right )B^4.
\end{align}
These expressions reduce to the Kerr-Newman values on setting $B=0$. As the geometry is not asymptotically flat, the definition of the Ernst-Wild black hole mass ${\mathcal M}$ is more subtle, but we will follow \cite{Astorino_2016} who observed that it can be consistently expressed in the form 
\begin{align}
    \mathcal M ^2 = \frac{\mathcal{S}}{4 \pi} + \frac{\mathcal{ Q}^2}{2} + \frac{\pi(\mathcal Q ^4+ 4 \mathcal J^2)}{4 \mathcal S},
\end{align}
which depends on $B$ implicitly through the other charges, and matches the Christodoulou-Ruffini mass as originally obtained for the Kerr-Newman black hole \cite{Christodoulou:1971pcn}. The black hole entropy ${\mathcal S}$ is given in terms of the area of the black hole horizon as $\mathcal S = A_+/4 = \pi \Lambda_0  (r_+^2 +a^2)$. 

In what follows, we will label Ernt-Wild solutions by their seed Kerr-Newman parameters $q, \, a$ and $M$, having in mind the above relations to the physical charges. Note that extremal black holes in the Ernst-Wild spacetime are characterized by the relation
\begin{equation}
    \label{e:extremalRelation}
    M = \sqrt{a^2 + q^2},
\end{equation}
where $r_H = M $ determines the radius of the degenerate event horizon. When $q=0$, there is no degenerate horizon and there is no additional restriction on the spin $a$, as can be seen in the first plot of Fig.~\ref{fig:extremalLimit}. However, when $q$ is nonzero (and we will consider the case of the Wald charge $-2 a M B$ and also the value $-a M B$), there is indeed a restriction on the parameters as shown in Fig.~\ref{fig:extremalLimit}. The location of the critical spin and magnetic field contour follows from solving Eq.~\ref{e:extremalRelation}, 
\begin{align}
    B_{c} = \begin{cases}
        \dfrac{\sqrt{M^2 -a^2}}{a M},  & \text{ for }  q=-a M B,\\
        & \\
        \dfrac{\sqrt{M^2 -a^2}}{2aM}, & \text{ for }   q=-2a M B.
    \end{cases}
\end{align}
To obtain a neutral black hole with conserved total charge $\mathcal{Q}=0$, using Eq.~\ref{e:QEW} the seed Kerr-Newman black hole charge must satisfy the condition, 
\begin{align}
    \mathcal{Q} = q\left(1- \frac{1}{4}q^2B^2 \right ) + 2 a M B = 0,
\end{align}
where in the limit $q B \ll 1$, the Wald charge \cite{Wald:1974np} is recovered, namely $q=-2 a M B$.

In general, the Ernst-Wild black hole is not strictly asymptotically Melvin due to the infinite extension of its ergoregion \cite{Gibbons:2013EW,Brenner_2021}. Consideration of the Killing vector $\xi_{\Omega} = \partial/\partial t + (\Omega/\Lambda_0)\, \partial/\partial_\phi$, with constant angular velocity $\Om$, shows that $\xi^\mu_\Omega\, \xi^\nu_\Omega\, g_{\mu\nu}$ will generally become large and positive, which means that there is an ergoregion near the rotation axis that extends to infinity, no matter the choice of $\Omega$.
However, this behaviour can be eliminated by choosing a particular value for the charge parameter in the solution, namely $q= - a M B$. Inputting this charge value into Eq.~\ref{e:QEW} corresponds to a total charge $\mathcal Q$ of 
\begin{align}
    \mathcal{Q}= a M B + \frac{1}{4} a^3 M^3 B^5.
\end{align}
In other words, when $q=- a M B$, the Killing vector $\xi_{\Omega}$ is timelike everywhere at large radius, and so the ergoregion is confined to a neighbourhood of the horizon. 
It is curious that in the small-$B$ limit this seed charge is precisely one-half that obtained by Wald \cite{Wald:1974np}.

\subsection{Apparent horizon}
For asymptotically flat geometries, the existence of an event horizon is a well-defined concept. 
That is, there exists a null hypersurface which is the boundary of the set of points which can be connected (by a causal curve) to null infinity \cite{Smarr:1973zz}. Determining its location requires knowing the global structure of the spacetime, which can be problematic when the geometry is non-asymptotically flat. Therefore, the existence of a black hole region in the Ernst-Wild spacetime will be justified by the identification of the apparent horizon, which is a local property of the spacetime. The surface geometry for a variety of black hole event horizons has been explored in numerous works; see, for instance, \cite{Smarr:1973zz,Wild:1980zz,WildKerns_PhysRevD.23.829,Kulkarni_PhysRevD.33.2780,Frolov_2006,Gibbons:2009qe,Annulli:2022ivr}.
In this section,  we simply present the apparent horizons for the Kerr and Ernst black holes in order to motivate the shape of the apparent horizon for the Ernst-Wild black hole, which was first investigated in \cite{Wild:1980zz,WildKerns_PhysRevD.23.829,Kulkarni_PhysRevD.33.2780} for the rotating and non-rotating Ernst-Wild spacetime, respectively. Specific details of the calculations are given in Appendix~\ref{app:apparenthorizon}.  

We confirm the findings of \cite{Smarr:1973zz} that as the rotation parameter increases, the apparent horizon of the Kerr black hole becomes oblate. In contrast, for the Ernst black hole, as the magnetic field strength increases the apparent horizon becomes prolate agreeing with \cite{Cunha:2018acu}. Since the Ernst-Wild black hole possesses both rotation and magnetic field, we can conclude that a black hole of this kind should exhibit both of these competing behaviours as the magnetic field and rotation both increase. This was shown in \cite{Annulli:2022ivr} for the charge values $q=0$ and $q=- aM B$. It is perhaps surprising that these two effects compete in this way and yield a more spherical horizon shape as both parameters are increased. Indeed, we will observe that this competition becomes even more apparent when we consider the black hole shadow.

\subsection{Weakly-magnetized Ernst-Wild}
\label{sec:WeaklyMagnetizedEW}
In order to characterize the strength of astrophysical magnetic fields, we define the scale $B_M = 1/M$,
corresponding to magnetic fields whose curvature effects are comparable to those at the horizon,
as in \cite{Brito:2014super}. After restoring physical units, this characteristic scale is given by
\begin{align}
    B_M \sim 2.4\, \times\, 10^{19} \left (\frac{M_{\odot}}{M} \right )\, \text{Gauss}.
\end{align}
Observationally, the largest magnetic fields that exist around compact objects are of the order {$ 10^{13}$--$10^{15}\, \text{Gauss}$} \cite{Olausen:2013bpa}. In natural units, this corresponds to $B/B_M \sim 10^{-6}$--$10^{-4}$. 
In order to explain the luminosity of some of the active galactic nuclei it is estimated that the magnetic field strength would be approximately $B \sim 10^{4}\, \text{Gauss} \sim 10^{-6} B_M$ for supermassive black holes with $M \sim 10^{9} M_\odot$ - which additionally must assume specifics of the interaction model between the black hole and the accretion disk \cite{piotrovich2010magneticfieldsblackholes}. 
Therefore, a perturbative expansion in the magnetic field is well justified astrophysically on the grounds that magnetic fields are not expected to be large compared to the mass of the black hole, i.e. $B\ll 1/M$, \cite{Frolov:2010mi,Frolov_2012,Brito:2014super}. 

The restriction to the regime $B\ll B_M$ leads to an important scale separation in the Ernst-Wild geometry, namely that the Melvin radius $r_M$, where the asymptotic Melvin-like geometry starts to dominate, is parametrically larger than the horizon 
\be
r_M \sim 1/B\gg r_+. 
\ee
This allows us to analyze the back-reaction of the magnetic field on the geometry in the inner region that is fully isolated from the pathologies of the non-asymptotically flat Melvin-like geometry at large radius. This renders the EW spacetime for $B\ll B_M$ a valuable model in the astrophysical context, for Kerr black holes immersed in an asymptotically uniform magnetic field. Moreover, this distinction also becomes apparent when considering perturbations. Although the EW geometry generally exhibits confining box-like boundary conditions \cite{Brito:2014super}, by expanding perturbatively to ${\cal O}(B^2)$ the far-field asymptotics are instead wave-like at spatial infinity, as in an asymptotically flat geometry \cite{Konoplya:2007yy,Becar:2022wcj,Taylor:2024duw}. At $\mathcal{O}(B^2)$, the modes are primarily sensitive to the localized impact of the magnetic field near the photon sphere.  We will make use of this feature in Section~\ref{sec:EWinPerturbativeRegime}.

\section{Light rings and equatorial geodesics}
\label{sec:Light rings and equatorial geodesics}
\begin{figure}
  \centering
\includegraphics[width=0.85\linewidth]{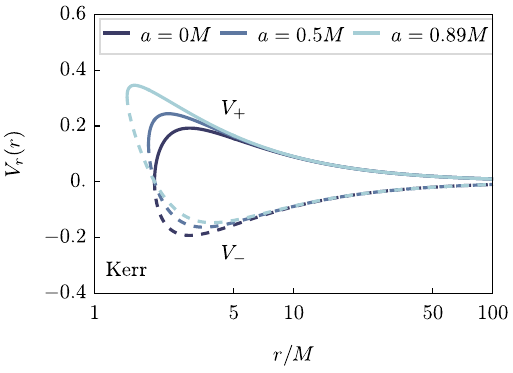}  
\vspace*{-0.1cm}
\caption{The potentials $V_+(r)$ and $V_-(r)$ for co-rotating equatorial orbits around the Kerr black hole with spin parameters $a=(0,0.5M, 0.89M)$. The region between the solid ($V_+$) and dashed ($V_-$) curves is not accessible to the motion of massless particles.}
\label{fig:KerrPotentials}
\end{figure}
\begin{figure*}
    \centering
    \includegraphics[width=\linewidth]{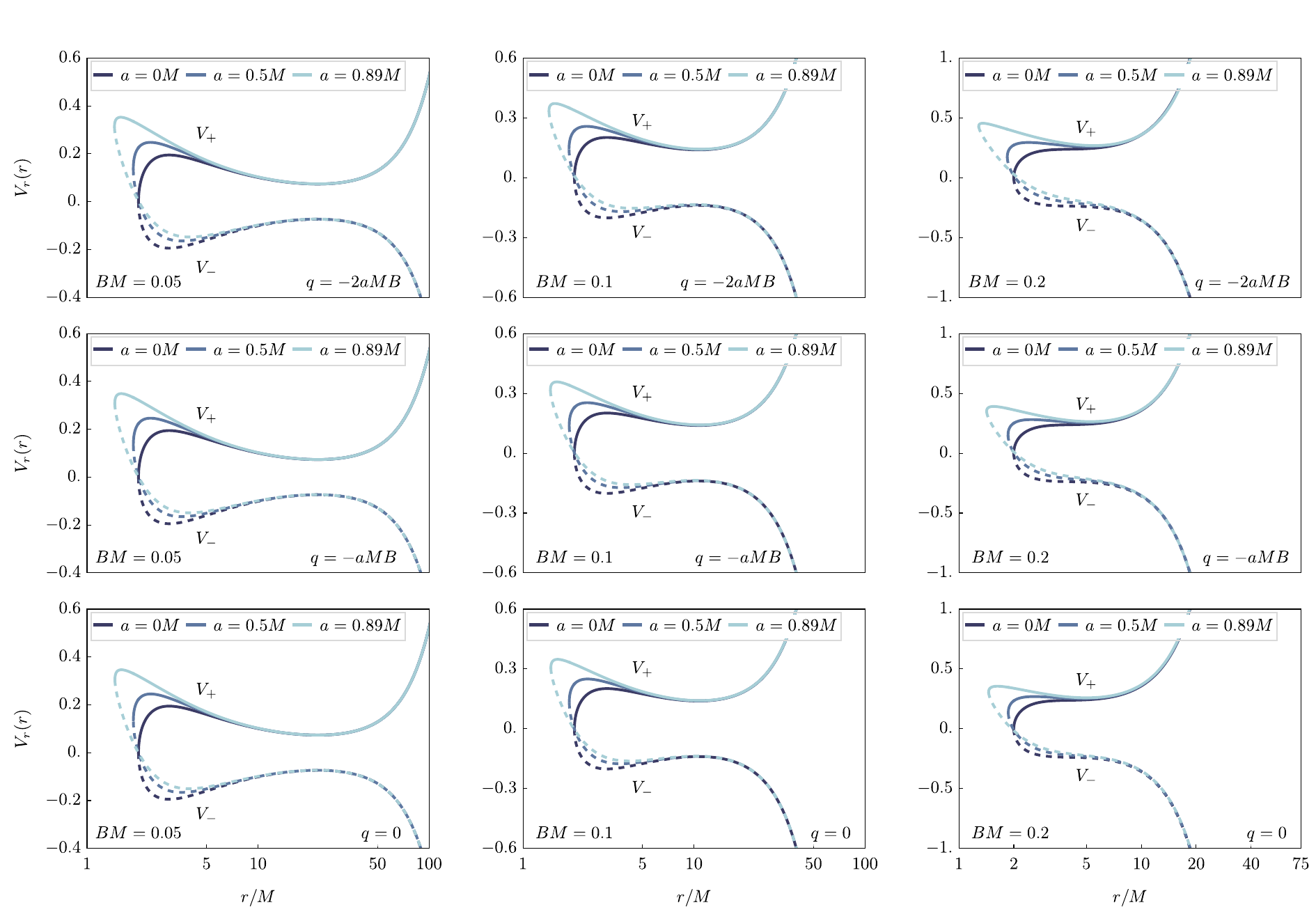}
    \caption{The potentials $V_+(r)$ and $V_-(r)$ for co-rotating (prograde) equatorial orbits around the Ernst-Wild black hole with spin parameters $a=(0,0.5M, 0.89M)$ for three values of the magnetic field $BM=(0.1,0.2,0.25)$ for $q=(0, -a M B, -2 a M B)$. The region between the solid ($V_+$) and dashed ($V_-$) curves is not accessible to the motion of massless particles.}
    \label{fig:Potentials_q=0}
\end{figure*}
\begin{figure*}
    \centering
    \includegraphics[width=\linewidth]{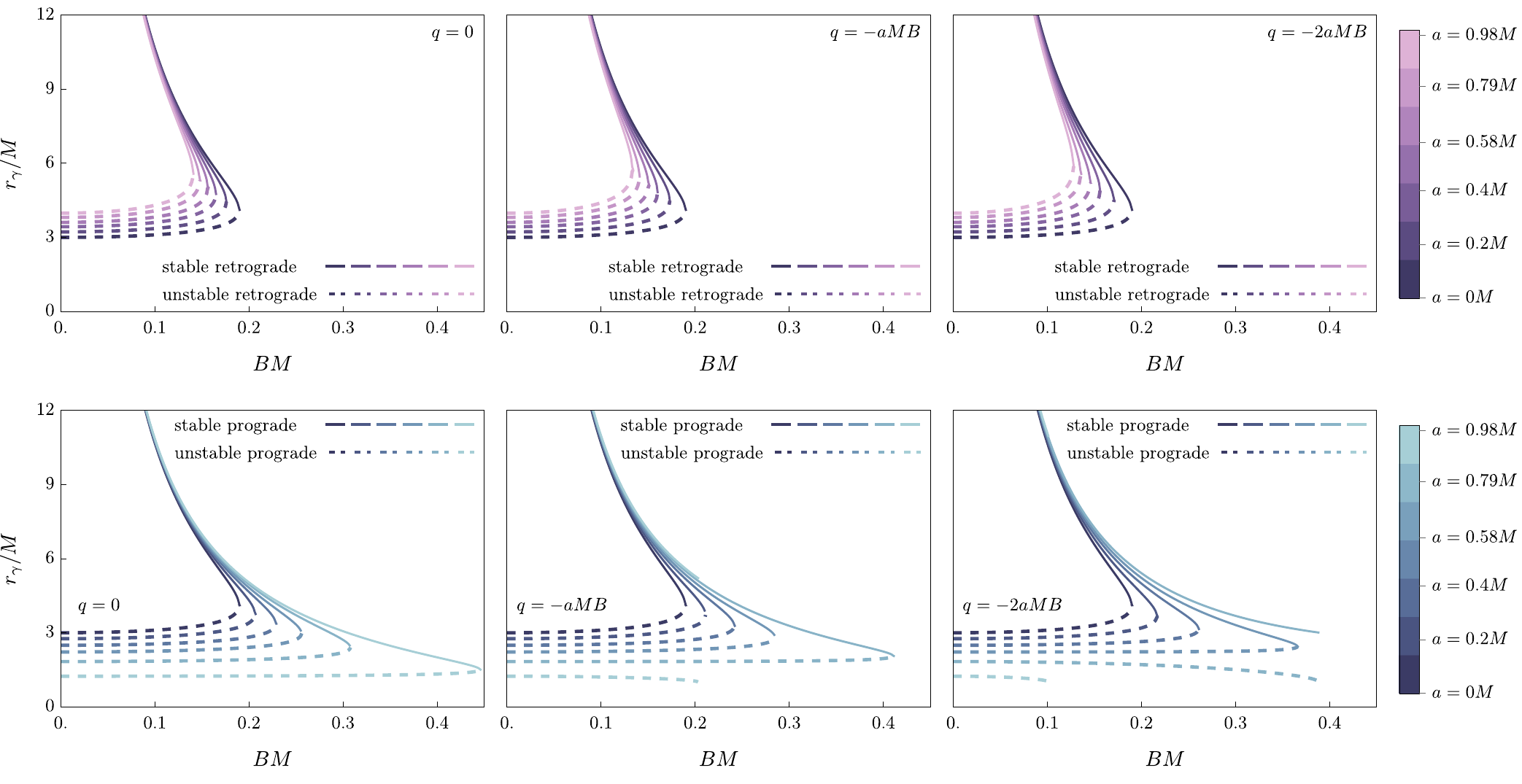}
    \caption{Retrograde (purple) and prograde (blue) stable and unstable dimensionless light ring radii $r_{\gamma}$ are shown for the seed Kerr-Newman charges $q=(0, \, -aMB,\,-2aMB)$ for six values of the spin $a$. For $q\neq 0$, as the magnetic field increases toward critical values, the unstable prograde orbits progressively turn inward toward the event horizon, eventually merging with it and causing the orbit to disappear. This behavior, along with additional constraints due to the extremality condition (see Fig.~\ref{fig:aCritbCrit}), distinguishes the $q\neq 0$ cases.}
    \label{fig:ProgradeRetrograde2}
\end{figure*}
Light rings are a special class of null geodesics, defined for spacetimes that possess at least two commuting Killing vectors, hereafter denoted $ \xi$ and  $\zeta$, with $[\xi, \zeta ] = 0$. These are associated with the stationarity and axial symmetry of the spacetime, respectively, and can be expressed in the coordinates of the spacetime $t, \phi$ as $\xi = \partial_t$ and $\zeta = \partial_\phi$. Any null vector tangent to a light ring is spanned by the vectors $(\xi,\, \zeta)$, which means geometrically that they are tied to these symmetries. Light rings can be classified according to their dynamical stability under perturbations. The existence of an unstable light ring permits light to encircle a black hole any number of times before either being scattered back to infinity or ultimately plunging into the black hole. Observationally they are also a crucial feature since these unstable light rings contribute to the boundary of the black hole shadow. However, in general, light rings are not necessarily connected to the shadow edge, particularly if multiple unstable light rings are present \cite{Cunha:2018acu}. In contrast, if stable light rings are perturbed, they can still revolve closely to the equilibrium trajectory. Although stable light rings do not appear in typical astrophysical scenarios, for instance the Kerr metric does not exhibit stable light rings, they are a feature present in more exotic geometries such as Proca stars \cite{Herdeiro:2021lwl}, Kerr black holes with bosonic hair \cite{Cunha:2015yba,Cunha:2016bpi,Cunha:2016bjh} and in particular black holes in asymptotically Melvin-like geometries \cite{Junior_2021}, the case explored in this work. 
\subsection{Magnetized spacetimes}
In this section, we will explore null geodesics in the Ernst-Wild geometry, focusing on the presence of light rings in the equatorial plane and the dependence of the orbit parameters on particular values for the charge and magnetic field. This topic has been explored in special cases in prior work. For example, in analyses neglecting back-reaction \cite{Zhong_2021,Lee_2023}, in the absence of a seed electric charge $q=0$ \cite{Wang_2021,Hou_2022,zhang2024energy,Jumaniyozov:2024hlg}, in the presence of a seed magnetic charge $p\neq 0$ \cite{yuan2024qed}, and in the absence of spin \cite{Junior_2021,He:2022opa,Shaymatov_2021,Shaymatov_2022}. There are also analyses in restricted regimes \cite{Zhong_2021,Lee_2023,yuan2024qed} and for charged particles \cite{Lee_2023,zhang2024energy,Al_Zahrani_2022,Santos:2024tlt}. We add to the literature by exploring two specific values of the seed charge parameter which are of interest for different reasons. The value $q=- a M B$ is of primarily theoretical interest since it leads to a compact ergoregion and exhibits Melvin asymptotics. 
The charge $q=- 2 aM B$ is chosen since it allows the solution to remain electrically neutral and, interestingly, incorporates the result of Wald \cite{Wald:1974np} for the charge induction in the small-$B$ limit.

We will begin by introducing the general form for a stationary and axisymmetric metric,
\begin{equation}
    \label{e:generalAxisymmetricStationaryMetric}
    ds^2 = g_{tt} dt^2 + g_{rr} dr^2 + g_{\theta \theta} d\theta^2+g_{\phi \phi} d\phi^2 +  2g_{t \phi} dt d\phi.
\end{equation}
With the EW geometry in mind, we consider the region beyond the outer horizon ($ r\geq r_+$ in Boyer-Lindquist coordinates). Null geodesics 
 with affine parameter $\lambda$ and tangent vector $u^\mu = \frac{d x^\mu}{d \lambda} \equiv \dot{x}^\mu$, satisfying  $u^\mu \nabla_\mu u^\nu = 0$, follow from the Lagrangian $\mathcal L (x^\mu, \dot x^\mu ) = \frac{1}{2} g_{\mu \nu} \dot{x}^\mu \dot{x}^\nu$.
 
For our purposes, it will be convenient to use a Hamiltonian formulation, with 
\begin{equation}
    \label{e:Hamiltonian}
    \mathcal H (x^\mu, p_\mu) = p_\mu \dot x^\mu - \mathcal{L}(x^\mu, \dot x^\mu) = \frac{1}{2} g^{\mu \nu}p_\mu p_\nu.
\end{equation}
with the conjugate momenta $p_\mu \equiv \partial \mathcal L / \partial \dot x^\mu = g_{\mu \nu} \dot x^\nu$. The geodesic equations are then equivalent to the Hamilton-Jacobi equations, 
\begin{align}
    \label{e:HJ_equations}
   \dot{x}^\mu = \frac{\partial \mathcal H}{\partial p_\mu}, \quad    \dot{p}_\mu = -\frac{\partial \mathcal{H}}{\partial x^\mu}.
\end{align}
The conserved quantities admitted by the general spacetime can be written as
\begin{equation}
    \label{e:conservedEandL}
    \begin{split}
        E \equiv -p_t = - \xi^\mu u_\mu &= - g_{tt} \dot x_t - g_{t \phi} \dot x_\phi,\\
        L \equiv p_\phi = \zeta^\mu u_\mu &= g_{t\phi} \dot x_t + g_{\phi \phi} \dot x_\phi,
    \end{split}
\end{equation}
where $E$ and $L$ are usually interpreted as the energy and angular momentum of the light ray measured by an observer at spatial infinity, but this definition relies on the metric being asymptotically flat. Although the Ernst-Wild geometry is stationary and axisymmetric, it is not asymptotically flat, which complicates the identification of a preferred observer and thus the physical identification of $E$ and $L$. 

 \begin{figure*}
    \centering
    \includegraphics[width=\linewidth]{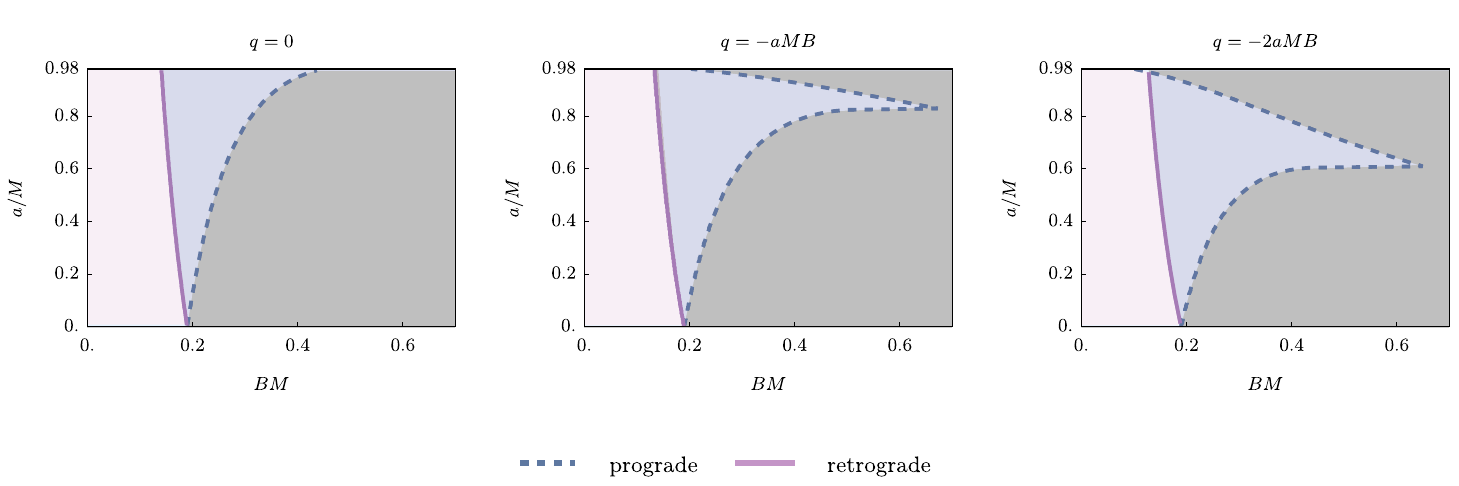}
    \caption{Critical values of the spin parameter $a$ and magnetic field $B$ for prograde (blue) and retrograde (purple) light ring radii $r_\gamma$. The left-most purple shaded region corresponds to the regime where both prograde and retrograde light rings exist, the middle blue shaded region is where only prograde light rings exist and the right-most dark gray shaded region is where prograde and retrograde light rings cease to exist. For $q \neq 0$, the shape of the allowed region is additionally constrained due to the Kerr-Newman extremal limit. Unlike Fig.~\ref{fig:ProgradeRetrograde2}, which plots $r_\gamma$ for a discrete number of spin values, here we plot the allowed parameter space in magnetic field and spin for the entire range to determine the values of $a_c$ and $B_c$.   }
    \label{fig:aCritbCrit}
\end{figure*}
Solving Eq.~\ref{e:conservedEandL} for $\dot{t}$ and $\dot{\phi}$ yields
\begin{align}
    \label{e:tdotphidot}
    \dot t = \frac{E g_{\phi \phi} + L g_{t \phi}}{g_{t \phi}^2 - g_{tt} g_{\phi \phi}}, \quad \dot \phi =- \frac{E g_{t \phi} + L g_{t t}}{g_{t \phi}^2 - g_{tt} g_{\phi \phi}}.
\end{align}
In the equatorial plane, with $\theta=\pi/2$ and $\dot \theta=0$, Eq.~\ref{e:Hamiltonian} simplifies and we can write
\begin{align}
    \mathcal H = g^{rr}p_r^2 - \frac{g_{\phi \phi} E^2 + 2 g_{t \phi} E L + g_{tt} L^2 }{g_{t \phi}^2 - g_{tt} g_{\phi \phi}} =0,
\end{align}
where we identify 
\begin{align}
    \label{e:rdotSquared}
    \dot{r}^2 = \frac{1}{g_{rr}}\left ( \frac{g_{\phi \phi} E^2 + 2 g_{t \phi} E L + g_{tt} L^2 }{g_{t \phi}^2 - g_{tt} g_{\phi \phi}}\right ).
\end{align}
After some algebra, Eq.~\ref{e:rdotSquared} can be written in the following useful form
\begin{align}
    \label{e:effectivePotVr}
    V_r(r) = \dot{r}^2= \frac{g_{\phi \phi}}{g_{rr}}(E- V_+)(E-V_-),
\end{align}
with
\begin{align}
    V_{\pm}= \frac{L}{g_{\phi \phi}} \left \{-g_{t \phi} \pm \sqrt{g_{t\phi}^2 - g_{tt}g_{\phi \phi}}\,\right \} .
\end{align}
When studying circular orbits in the equatorial plane, it must be that light rings also satisfy
\begin{equation}
    \label{e:potentialandDerivVanish}
        V_r(r) = 0, \quad 
        \frac{\partial V_r(r)}{\partial r} = 0.
\end{equation}
Additionally, the light ring is classified as an unstable circular photon orbit if
\begin{equation}
    \frac{\partial^2 V_r(r)}{\partial r^2} < 0, 
\end{equation}
and the light ring is classified as a stable circular photon orbit if
\begin{equation}
    \frac{\partial^2 V_r(r)}{\partial r^2} > 0.
\end{equation}
Since $\dot{r}^2$ must be positive, massless particles must move on geodesics with constant of motion $E$ satisfying the following inequalities
\begin{equation}
    E < V_- \quad \text{or} \quad E> V_+,
\end{equation}
meaning that the region $V_- < E < V_+$, is not an allowed region. In Fig.~\ref{fig:KerrPotentials}, we plot the Kerr result for three values of the spin parameter for comparison to the Ernst-Wild result shown in 
Fig.~\ref{fig:Potentials_q=0}. Here we plot the potentials $V_+$ and $V_-$ for $q=-2 a M B$, $q=- a M B$ and $q=0$ for the same values of the spin parameter. 

The prograde and retrograde light ring orbits for the Ernst-Wild spacetime can be seen in Fig.~\ref{fig:ProgradeRetrograde2} 
where we have plotted the dimensionless light ring radius $r_{\gamma}/M$ for the values $q=\{0,- a M B, -2a M B\}$. As expected, the $q=0$ plots match the results obtained in \cite{Wang_2021}, but the added charge parameter changes the characteristics of the prograde light ring orbits as the spin and magnetic field increase.
Indeed, since the Ernst-Wild black holes must satisfy the extremal condition, namely $\sqrt{a^2 + q^2}\leq M$, as can be seen in Fig.~\ref{fig:extremalLimit}, there will be additional constraints on the spin and magnetic field values. Additionally, for $q\neq 0$, as $a\rightarrow a_c$, the prograde orbits will begin to turn towards the event horizon for critical values of the spin, as inferred from Fig.~\ref{fig:extremalLimit}. These inturned orbits stay close to the location of the event horizon until they reach it, and at this point the orbit disappears; that is, when the light ring coincides with the event horizon the orbit disappears. 

The general characteristics of the orbits are summarized below: 
\begin{itemize}
    \item As the magnetic field $B$ increases, the radius $r_{\gamma}$ of all unstable retrograde orbits, and most unstable prograde orbits, increases, whereas $r_{\gamma}$ decreases for stable prograde and retrograde orbits.
    \item As the magnetic field $B\rightarrow 0$, 
    the radius of stable light rings $r_{\gamma} \rightarrow \infty$, recovering the Kerr result.
    \item As the spin $a$ increases, the radius $r_{\gamma}$ decreases for stable retrograde orbits and most unstable prograde orbits, while it increases for stable prograde orbits and unstable retrograde orbits.
    \item For each fixed rotation parameter $a$, the radius $r_{\gamma}$ disappears at a critical magnetic field $B_c$.
\end{itemize}
These features match the analysis carried out for $q=0$ in \cite{Wang_2021}. However, when the charge $q \neq 0$, we observe the following additional behaviors:
\begin{itemize} 
    \item For unstable prograde orbits with spin parameters above a critical value $a_c$, the light ring radius $r_\gamma$ asymptotes to the outer event horizon $r_+$ near extremality. This feature is illustrated in Fig.~\ref{fig:aCritbCrit}, which depicts the allowed parameter space for the magnetic field and spin. 
    \item Retrograde orbits, while largely unaffected, approach similar but numerically different critical values. 
\end{itemize}

Since the Hamilton-Jacobi equations given by Eq.~\ref{e:HJ_equations}, are not in general separable as shown explicitly in \cite{Bezdekova:2022gib}, we will use numerical methods to explore the light ring and photon shell more generally. In order to visualize the effects of the strong gravitational fields, the light trajectories need to be tracked in order to determine which eventually reach the observer and those which enter the black hole. Placing an observer at the origin and evolving the null geodesics that reach the observer backwards in time allows the reconstruction of the bright source from where the light originates. This technique is the standard method utilized for simulating black hole shadows and is referred to as the backwards ray tracing technique and will be covered in more detail in the next section.

\section{Black hole shadows and backwards ray-tracing}
\label{sec:RayTracing}
\begin{figure}
    \centering
    \includegraphics[width=0.6\linewidth]{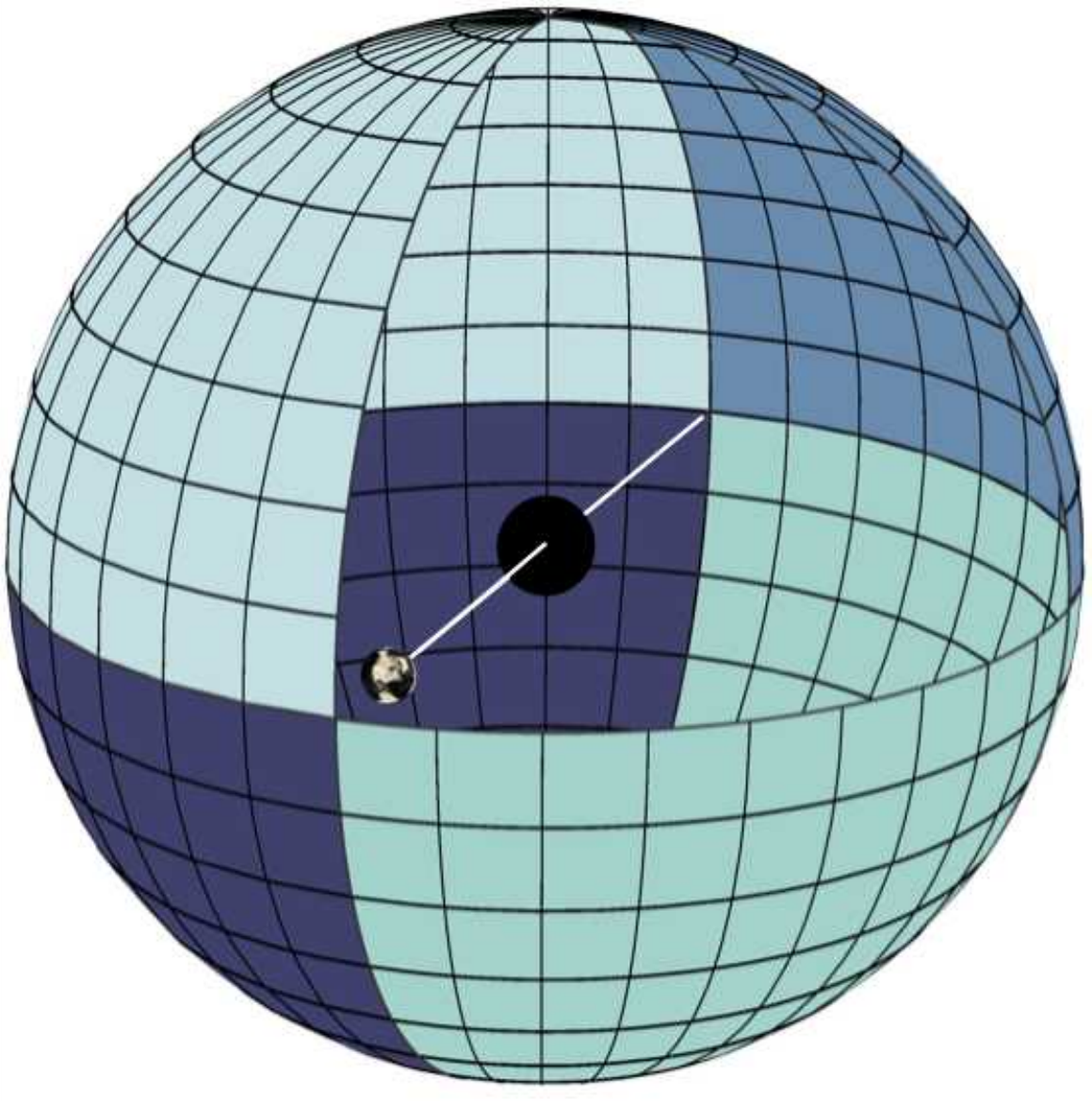}
    \caption{Schematic of the celestial sphere used in the ray tracing images where we have removed a section of this to see the observer and the black hole, where the white line represents the direction of radial observation. Each quadrant of the sphere is coloured a different shade to distinguish where the light rays originate, adapted from \cite{VelasquezCadavid:2022pex}.}
    \label{fig:celestial_sphere}
\end{figure}

Turning to gravitational lensing in the Ernst-Wild spacetime, we now proceed to visualize the black hole geometry for non-zero charge $q$. 
The backward ray-tracing method \cite{Bohn:2014xxa} relies on placing the observer at some distance away from the black hole, within a brightly lit celestial sphere, and tracing the light that reaches the observer backward in the time-parameter to determine from where it originated. The black hole is placed at the  centre of this celestial sphere of size $r_{cs}$, with four quadrants coloured differently along with a dark gray mesh with constant longitude and latitude lines separated by $\pi/18$ radians, as can be seen in Fig.~\ref{fig:celestial_sphere}.
\subsection{Backwards ray tracing}
\begin{figure}
\begin{subfigure}{.23\textwidth}
  \centering
  \includegraphics[width=\linewidth]{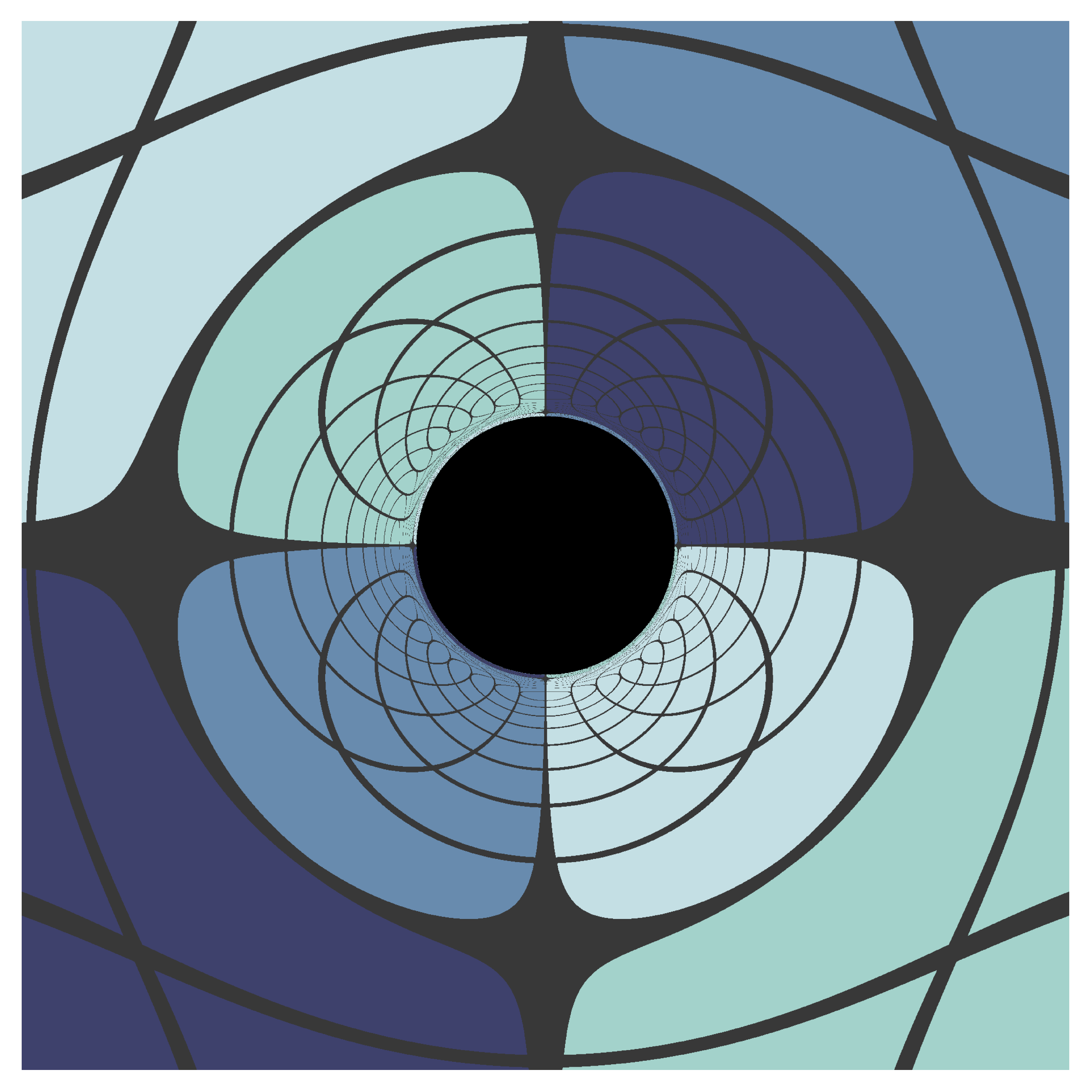}  
\end{subfigure}
\begin{subfigure}{.23\textwidth}
  \centering
  \includegraphics[width=\linewidth]{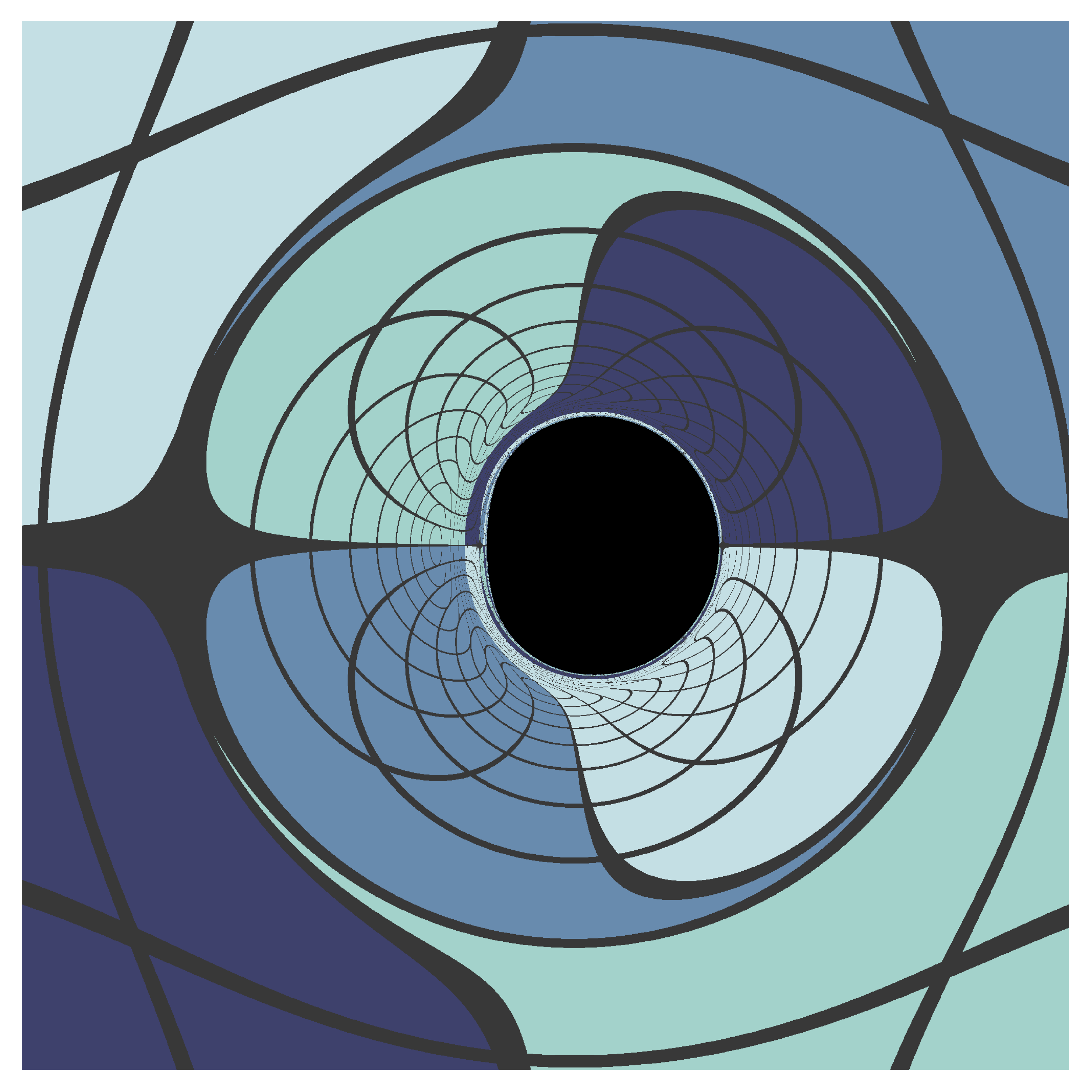}  
\end{subfigure}
\vspace*{-0.25cm}
\caption{The left image displays the Schwarzschild black hole and the right image displays the near-extremal Kerr ($a=0.98M$) black hole using the stereographic projection with $r_p = 50M,\, r_{cs} =25\,000 M$.}
\label{fig:raytracingStereographic}
\end{figure}
\begin{figure*}
    \centering
    \includegraphics[width=\linewidth]{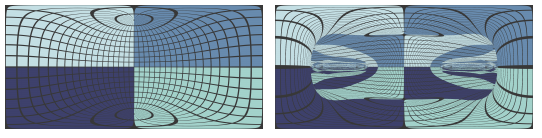}
    \caption{Minkowski spacetime (left) and Melvin spacetime (right) using the equirectangular projection to produce an anisotropic view. Minkowski spacetime: observer is placed at $r_p=1$ with the celestial sphere located at $r_{cs}=2.5$. Melvin spacetime: observer is placed at $r_p=1/B$ with the celestial sphere located at $r_{cs}=2.5/B$.}
    \label{fig:melvinMinkowski}
\end{figure*}
The four-momentum for each null geodesic, measured by the observer in the local (tetrad) frame, must first be determined. The observer's basis vectors in a local inertial frame $ \ehat{\nu}^\mu = ( \ehat{t}, \ehat{r}, \ehat{\theta}, \ehat{\phi})$, can be expanded in terms of the coordinate basis vectors $(\bl{t}, \bl{r}, \bl{\theta}, \bl{\phi})$, associated with the Boyer-Lindquist coordinates. One possible choice for this observer basis is
\begin{equation}
    \begin{split}
        \ehat{t} &= \varsigma \bl{t} +\gamma \bl{\phi},\\
        \ehat{r} &= \frac{1}{\sqrt{\gmunu{rr}}}\bl{r},\\
        \ehat{\theta} &= \frac{1}{\sqrt{\gmunu{\theta \theta}}} \bl{\theta},\\
        \ehat{\phi} &= \frac{1}{\sqrt{\gmunu{\phi \phi}}} \bl{\phi},
    \end{split}   
\end{equation}
where 
\begin{equation}
        \varsigma = \sqrt{\frac{g_{\phi \phi}}{g_{t \phi}^2 - g_{tt}g_{\phi \phi}}},\quad 
        \gamma = -\frac{g_{t \phi}}{g_{\phi \phi}}\sqrt{\frac{g_{\phi \phi}}{g_{t \phi}^2 - g_{tt}g_{\phi \phi}}}.   
\end{equation}
The normalization conditions for these vectors are defined by the scalar products, 
\begin{equation}
\begin{split}
    \ehat{r} \cdot \ehat{r}= \ehat{\theta} \cdot \ehat{\theta} = \ehat{\phi}\cdot \ehat{\phi} &= 1,\\
    \ehat{t} \cdot \ehat{t} &=-1,\\
    \ehat{t} \cdot \ehat{\phi} &=0.
\end{split}
\end{equation}
Locally, the observer perceives Minkowski spacetime and for this reason this observer is commonly called the zero-angular-momentum observer (ZAMO).
The components of the four-momentum of the photon measured in this frame are given by 
\begin{equation}
\label{e:photonZAMO}
    \begin{split}
        p^{(t)} &=- e^{\mu}_{(t)}p_\mu =  \varsigma E - \gamma L,\\
        p^{(r)} &= \hat e^\mu_{(r)}p_\mu  =\frac{1}{\sqrt{g_{rr}}}p_r,\\
        p^{(\theta)} &= \hat e^\mu_{(\theta)}p_\mu  =\frac{1}{\sqrt{g_{\theta \theta}}}p_\theta,\\
        p^{(\phi)} &= \hat e^\mu_{(\phi)}p_\mu  =\frac{1}{\sqrt{g_{\phi \phi}}}L.
    \end{split}    
\end{equation}
The photon's linear three-momentum $\vec p$ in the observer frame has the components
\begin{align}
    \vec{p} = (p^{(r)}, p^{(\theta)}, p^{(\phi)}),
\end{align}
which are parametrized in  terms of the celestial coordinates $(\alpha, \beta)$ as 
\begin{equation}
\label{e:photonCelestial}
\begin{split}
    p^{(r)} &= |\vec{p}| \cos\alpha \cos \beta,\\
    p^{(\theta)} &= |\vec{p}| \sin\alpha,\\
    p^{(\phi)} &= |\vec{p}| \cos\alpha \sin \beta.  
\end{split}
\end{equation}
Using Eq.~\ref{e:photonZAMO} and Eq.~\ref{e:photonCelestial} we can identify the following, 
\begin{equation}
\begin{split}
    E &= \rthO{|\vec{p}| \left ( \frac{1 + \gamma \sqrt{g_{\phi\phi}} \cos \alpha \sin \beta}{\varsigma}\right )},\\
    \dot r &= \rthO{|\vec{p}| \sqrt{g_{rr}}\cos\alpha \cos \beta},\\
    \dot \theta &= \rthO{|\vec{p}| \sqrt{g_{\theta \theta}}\sin \alpha},\\
    L &= \rthO{|\vec{p}| \sqrt{g_{\phi \phi}}\cos\alpha \sin \beta},   
\end{split}      
\end{equation}
where $r_o$ and $\theta_o$ are the observer's coordinates. We set the value of $|\vec p|=E =1 $ to unity without loss of generality, as it's variation simply rescales the affine parameter $\lambda$ of the geodesic trajectory. 
 
The solid angle that an object occupies in the observer's local sky depends on the distance between the object and the observer and is more subtle in a curved spacetime. 
Therefore, a better notion of distance, the perimetral (or circumferential) radius,  will be introduced following \cite{Cunha:2015yba,Cunha:2016bpi}.
We will say that black holes can be observed under comparable conditions provided that the perimetral radius, 
\begin{align}
    \label{e:rp}
    r_p =\left .\sqrt{g_{\phi \phi}}\right |_{\theta = \pi/2},
\end{align}
is the same in both geometries \cite{Cunha:2018acu,Junior_2021}, rather than the radial coordinate $r$. 
Therefore, Ernst-Wild black holes and Melvin universes will be said to be observed under comparable conditions provided that the perimetral radius given by 
\begin{align}
    r_p^2 & = \frac{16 \Lambda_0^2 \left(r^4+a^2 \left(r (2 M+r)-q^2\right)\right)}{ \left( (2 M+r)a^2B^2+4 a B q+r \left(4+B^2 r^2\right)\right)^2}
\end{align}
is the same in both cases. In the limit $a\rightarrow0$ and $q\rightarrow0$, we confirm the result from \cite{Junior_2021} for the Ernst black hole with 
\begin{align}
    \label{e:rpErnst}
    r_p= \frac{4 r}{4+B^2 r^2}.
\end{align} 
\begin{figure}[t]
    \centering
    \includegraphics{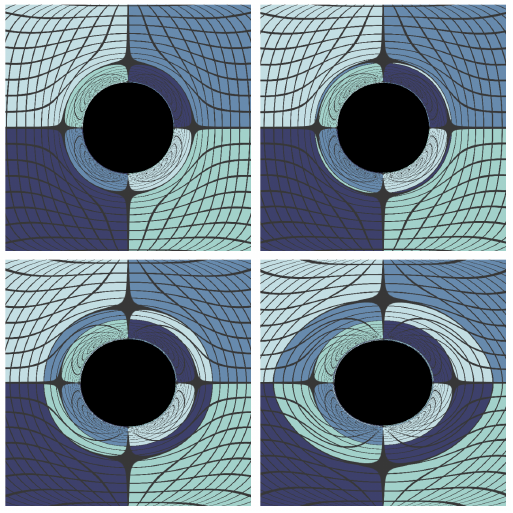}
    \caption{Ernst black hole with $q=0, \, a=0$, top row: $BM = (0, 0.02)$ and bottom row $BM = (0.04, 0.05)$ for $r_p=10M$, $r_{cs} =25M$. The top left corner with $BM=0$ corresponds to a comparable Schwarzschild black hole.}
    \label{fig:raytracingBnonzero}
\end{figure}
\begin{figure}
    \centering
    \includegraphics{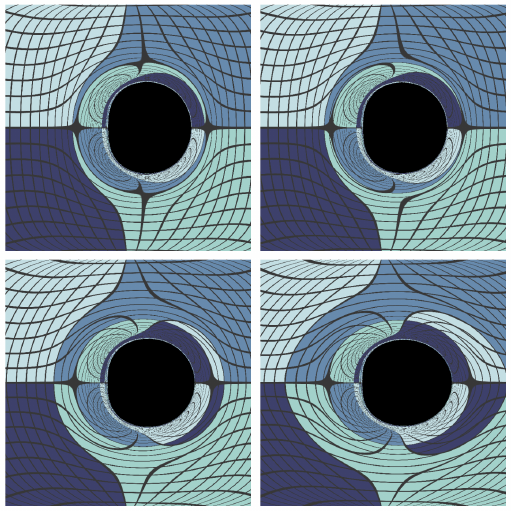}
    \caption{Ernst-Wild black hole with $q=-2 a M B$, $a=0.98M$, top row $BM = (0, 0.02)$ and bottom row $BM = (0.04, 0.05)$ for $r_p=10M$, $r_{cs} =25M$. The top left corner with $BM=0$ corresponds to a comparable Kerr black hole.}
    \label{fig:raytracingLargeSpin}
\end{figure}
\begin{figure*}
    \centering
    \includegraphics[width=\linewidth]{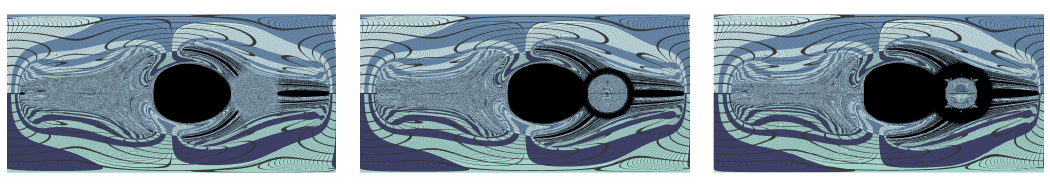}
    \caption{Panoramic shadow for the Ernst-Wild black hole with charges $q=(0,\, -a M B, \, -2 a M B)$ from left to right for $a=0.98M$ and $BM=0.1$. The observer is located at $r_p=7M$ and the celestial sphere is located at $r_{cs}=45M$. }
    \label{fig:panoramicAllqs}
\end{figure*}
Solving Eq.~\ref{e:rpErnst} if $B\neq 0$, the perimetral radius has a maximum value given by 
\begin{align}
    r_p^{\text{max}} = \frac{1}{B},
\end{align}
with a corresponding value of the radial coordinate 
\begin{align}
    r = \frac{2}{B},
\end{align}
which is characteristic of Melvin universes. 
Since $r\in (0, \infty)$, for Ernst-Wild spacetimes with $B\neq0$, there will be generically two radial coordinates $r$ for the same perimetral radius $r_p$. The observer location is chosen to be (of the two) the one with the smallest $r$-value. 

The camera model used by the observer will determine the mapping between the flat image plane and the observer's spherical sky. Clearly, the resulting black hole image will depend on the chosen camera model used within the ray-tracing algorithm since different models will differ in how a viewing direction is mapped onto the image plane. For asymptotically flat geometries, interesting gravitational lensing occurs near the centre of the image. Here, it is convenient to use a smaller field of view within the stereographic projection (also known as the fisheye camera model), and it has been used extensively in black hole imaging; see, for instance, \cite{Bohn:2014xxa,Hu:2020usx,Zhong_2021,yuan2024qed,Guo:2024mij}. 
Resulting images using this projection are displayed in Fig.~\ref{fig:raytracingStereographic} for the Schwarzschild and near-extremal Kerr black hole. This projection is particularly useful in asymptotically flat geometries, since it is advantageous in this projection to set the observer a very large distance away in order to capture more of the gravitational lensing structure. 
For non-asymptotically flat geometries, like the Melvin spacetime, non-trivial features can occur panoramically, requiring the use of the equirectangular projection; see  \cite{Cunha:2015yba,Cunha:2016bpi,Cunha:2018acu,Junior_2021} for more details.  

The equirectangular projection simply maps the spherical $\phi$ coordinate to the $y-$axis and the spherical polar coordinate $\theta$ to the $x-$axis, and is implemented as follows. In panoramic images, the celestial coordinates range from $\alpha \in\left [ -\pi/2,\, \pi/2 \right ]$ and $\beta \in\left [ -\pi,\, \pi \right ]$, reflecting a larger horizontal range than vertical. 
For each pixel in the final image, the initial conditions are determined by the celestial coordinates - therefore to project a given viewing direction $(\alpha,\, \beta)$ onto a point $(x,y)$ in the image plane we use
\begin{equation}
        \alpha  = y\, \text{fov}_y, \quad 
        \beta = x \, \text{fov}_x,
\end{equation}
where $\text{fov}_x$ and $\text{fov}_y$ correspond to the observer's field of view (fov) in $x$ and $y$ respectively.
In both projection models, the image coordinates can then be written in terms of pixel coordinates as
\begin{align}
    x &= \frac{1}{n_x}\left (i - \frac{1+n_x}{2} \right ),\\
    y &=\frac{1}{n_y}\left (j  - \frac{1+n_y}{2} \right ),
\end{align}
where $n_x$ is the number of pixels in the $x$--direction and $n_y$ is the number of pixels in the $y$--direction and $(i, j)$ label the pixel coordinates beginning from the bottom left hand corner at $(1,1)$ to the top right hand corner at $(n_x, n_y)$ of the resulting image.
In order to numerically integrate the geodesic equations, the initial conditions for the light ray
must be supplied: $\{t_{o},r_{o}, \theta_{o}, \phi_{o} \}$ and $\{p_{t_{o}},p_{r_{o}},p_{\theta_{o}},p_{\phi_{o}}\}$ at $\lambda=\lambda_o$ for a chosen field of view $(\text{fov}_x, \, \text{fov}_y)$ and an image size of $n_x\times n_{y} $.
If the light ray reaches the celestial sphere at $r_{cs}$, the integration stops, and one of the four colours in Fig.~\ref{fig:celestial_sphere} is assigned depending on the sector where light ray hits the sphere. Alternatively, if the light ray reaches the event horizon (within a small tolerance of $r_+\times10^{-5}$), the photon is captured, and the colour is recorded as black. This process is repeated until all pixels are assigned to their respective colours. 

In Fig.~\ref{fig:melvinMinkowski} we show the difference between the Minkowski spacetime and the Melvin spacetime. Since the magnetic field strength is the only scale in the Melvin geometry, all dimensionful quantities are normalized by $B$. For the Minkowski image, the observer is placed at $r_p = 1=r_o$ with the celestial sphere at $r_{cs}=2.5$. In the Melvin image, the observer is located at $r_p=1/B$ ($r_o=2/B$) with the celestial sphere at $r_{cs}=2.5/B$. This plot matches the one obtained in \cite{Junior_2021}. 

In Fig.~\ref{fig:raytracingBnonzero} we show ray tracing images of the Ernst black hole alongside the Schwarzschild black hole, that is $BM = (0,\,0.02,\,0.04,\, 0.05)$. Compared with the Schwarzschild figure in the upper left quadrant, the effect of the magnetic field can be seen with the shadow becoming increasingly oblate. These results match the images first shown in \cite{Junior_2021} for weak magnetic fields. In Fig.~\ref{fig:raytracingLargeSpin} we show ray tracing images of the Ernst-Wild black hole (with $q=-2 a M B$ ) alongside a comparable Kerr black hole, with  $a=0.98M$ for $BM = (0,\,0.02,\,0.04,\, 0.05)$. Compared to the reference Kerr image in the upper left corner, the shadow once again becomes increasingly oblate and deformed in a similar way as the Ernst black hole. In these images, the observer is placed at $r_o = 10 M$ with a celestial sphere at $r_{cs}=25 M$ and an equal field of view of $5\pi/6$.\footnote{We note that the images in Fig.~\ref{fig:raytracingLargeSpin} do not match those in the published version of \cite{Wang_2021}. However, the weakly-magnetized images in \cite{Wang_2021} should be taken with some care as those images seem to describe a configuration with unequal $r_p$ and where the observer is located outside the Melvin radius (above magnetic field strength $BM =0.03$).}

For clarity, we note that we have chosen to compare black holes with the same Kerr-Newman seed parameters, $(q, \, a, \, M)$ while varying $B$ in these images. Another reasonable alternative would have been to keep the total external charges $(\mathcal{Q,\, J,\, M})$ in (\ref{e:QEW}) fixed instead. For the non-rotating Ernst images this change has no effect, while the impact for rotating  Ernst-Wild solutions is also quite small for weakly-magnetized systems. For example, using the parameters in the bottom right plot of Fig.~\ref{fig:raytracingLargeSpin}, that is, $a=0.98$, $B=0.05$ and $M=1$, the difference between the physical EW charges $(\mathcal{Q,\, J,\, M})$ and the Kerr-Newman conserved charges $(q,\, aM,\, M)$ is $\sim 0.003 \%$. Thus the choice of which parameters are held fixed does not significantly impact the ray tracing results. 

Although our focus is on weakly-magnetized systems and the perturbative deviation of the shadow from the Kerr expectations, for completeness in  Fig.~\ref{fig:panoramicAllqs}, we display the ray tracing images and corresponding shadow for the Ernst-Wild black hole for a particular value of the spin and magnetic field resulting in a panoramic shadow, that is $a=0.98M $ and $BM=0.1$. In these images, we place the observer at $r_p=7M$, with a celestial sphere at $r_{cs}=45M$ and render the full field of view. Indeed, here the image for $q=0$ matches the one obtained in \cite{Wang_2021} for a comparable spin and magnetic field.

\subsection{Quantifying the shadow deformation numerically}
\begin{figure*}[t]
    \centering
    \includegraphics[width=\linewidth]{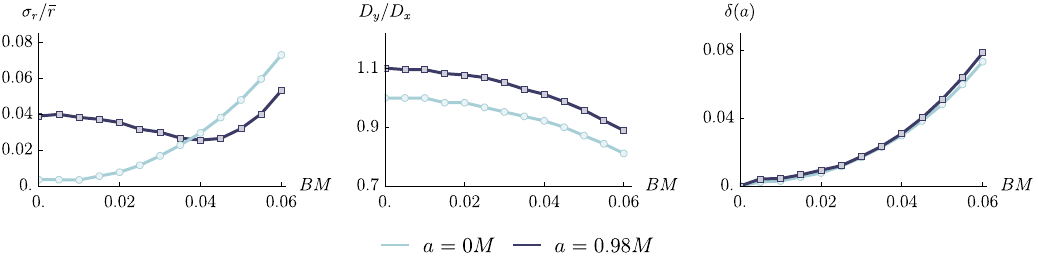}
    \caption{Shadow deformation parameters for the Ernst and Ernst-Wild black hole corresponding to spin values of $a=0M$ and $a=0.98M$ respectively. Each point on the curve represents the value computed from a ray tracing image in increments of $\delta B =0.005/M$.}
    \label{fig:shadowDeform2}
\end{figure*}
In order to quantify the impact of the magnetic field on the shadow numerically, we use the procedure first outlined in \cite{Johannsen:2013szh} with the notation developed in \cite{Cunha:2015yba} (and applied to QED effects of black holes, including presence of magnetic field in \cite{Zhong_2021}). Following \cite{Cunha:2015yba}, we introduce six parameters $\{D_C,\, D_x,\, D_y,\, \bar r,\, \sigma_r,\, \delta \}$, which will aid in quantifying the shadow deformation. 

The origin of the Cartesian coordinate system $O$ points at the centre of the celestial sphere where each of the four colours meet.  The shadow's centre, $C$, can be defined in terms of the minimum and maximum  ordinates of the shadow's edge, that is
\begin{equation}
    \begin{split}
        x_C &\equiv (x_{\text{max}}+x_{\text{min}})/2,\\
        y_C &\equiv (y_{\text{max}}+y_{\text{min}})/2.
    \end{split}
\end{equation}
Since there exhibits reflection symmetry on the equatorial plane, we can set $y_C =0$. The displacement $D_C = |x_C|$ measures the difference between the centre of the shadow $C$ and the origin $O$. We note that $D_C=0$ when the spin is zero. The width and height of the shadow can be obtained through
\begin{equation}
    \begin{split}
        D_x &\equiv x_{\text{max}} - x_{\text{min}},\\
        D_y &\equiv y_{\text{max}} - y_{\text{min}},
    \end{split}
\end{equation}
respectively. An arbitrary point $P$ on the edge of the shadow from $C$ is located at 
\begin{equation}
    r \equiv \left ({y_{P}}^2 + (x_P-x_C)^2 \right )^{1/2}.
\end{equation}
If we denote $\alpha$ as the angle between the line connecting $C$ and $P$ and the $x$-axis, the average radius can be written as 
\begin{equation}
    \bar r \equiv \frac{1}{2\pi}\int_0^{2\pi} d\alpha\, r(\alpha),
\end{equation}
and the deviation from sphericity is 
\begin{equation}
    \sigma_r \equiv  \sqrt{\frac{1}{2\pi} \int_{0}^{2\pi}d\alpha\, \left (r(\alpha) - \bar r \right )^2}.
\end{equation}
Finally, the relative deviation from a comparable Kerr black hole (or Schwarzschild if $a=0$) is given by 
\begin{equation}
    \delta(a) \equiv  \sqrt{\frac{1}{2\pi} \int_{0}^{2\pi}d\alpha\, \left (\frac{r(\alpha)-r_{\text{Kerr}}(\alpha,a)}{r_{\text{Kerr}}(\alpha,a)} \right )^2}.
\end{equation}

In Fig.~\ref{fig:shadowDeform2} we display the ratio of the deviation from sphericity and average radius $\sigma_r/\bar r$, the ratio of the height to width of the shadow $D_y/D_x$ and finally the deviation from a comparable asymptotically flat black hole $\delta (a)$ for the spin values $a=0$ and $a=0.98M$. For each curve, we computed a collection of images for magnetic field values equally spaced between $BM \in [0, 0.06]$ in increments of $\delta B = 0.005/M$, keeping the spin fixed.
It is interesting that in the figure depicting the deviation from sphericity, $\sigma_r$, that the Ernst-Wild result displays the behaviour motivated by the consideration of the apparent horizon and discussed further in Appendix~\ref{app:apparenthorizon}. As the magnetic field increases for $a=0.98M$, the deviation from sphericity initially increases but then for $BM\sim 0.05 $ begins to decrease before reaching the magnetic field value of $BM \sim 0.04$ where the deviation from sphericity once again increases and matches the qualitative form of the Ernst geometry. 
Perhaps unsurprisingly, the ratio of the shadow height to width decreased in both cases as the magnetic field increased, corresponding to the increasingly oblate nature of the shadow seen in both Fig.~\ref{fig:raytracingBnonzero} and Fig.~\ref{fig:raytracingLargeSpin}.

\section{Slow rotation and weak magnetization} \label{sec:EWinPerturbativeRegime}

This section considers perturbations of the Ernst-Wild geometry and its light ring structure. This is not straightforward in general as the system of perturbation equations is not separable for EW. Nonetheless, progress can be made by expanding perturbatively in $B$, which is well-motivated astrophysically, and also by considering small spin. This renders the perturbation equations tractable, and we will explore the extension of an intriguing geometric correspondence observed for slowly-rotating Kerr solutions \cite{Ferrari:1984zz,Yang:2012he}. Namely, the relation between high-frequency quasinormal modes and properties of the light ring null geodesics (in the eikonal regime). In particular, the real part of the QNM frequency corresponds to the light ring orbit, while the imaginary part corresponds to the Lyapunov exponent controlling the growth of perturbations away from the unstable light ring. We find that a perturbative expansion in $a$ and $B$ allows this correspondence to be extended to magnetized EW black holes with finite $B$.

To build up the components of this correspondence, in the next subsection, we first introduce the perturbative scaling regime to be used in computing the quasinormal modes. The Klein-Gordon equation for scalar perturbations then becomes separable, and we compute the high-$l$ QNM using  the sixth-order WKB method. The results are found to agree well with those obtained using the continued fraction technique \cite{Taylor:2024duw}, even for relatively low $l$. Unstable light ring perturbations in the eikonal regime are then considered utilizing the methodology described in \cite{Cardoso:2008bp} allowing us to identify the leading $B$-dependence of the Lyapunov exponents. In what follows, to be concrete we set the charge $q\equiv-2 a M B$ to ensure vanishing electric charge at infinity, and for the remainder of the paper, the term weakly-magnetized Ernst-Wild will also imply the limit of slow-rotation. In the plots of this next section,  we use the same colour scheme as Section~\ref{sec:Light rings and equatorial geodesics} where blue denotes prograde orbits and purple denotes retrograde orbits.

\subsection{Perturbative expansion in spin and magnetic field }
The analysis of linearized perturbations of the Ernst-Wild geometry was carried out in \cite{Brito:2014super} in the context of black hole superradiance. Considering only a perturbative expansion in the magnetic field $B$, while still simpler than considering the full geometry, still does not allow full separation of variables, and moreover it introduces couplings that link modes of different multipole number $\ell$. 
Nevertheless, if, in addition to considering small magnetic fields, we make a second simplifying assumption: that the black hole is slowly-rotating, the perturbation equations decouple completely, and standard techniques can be applied. 
To leading order in $a$ and $B^2$, the magnetic field effectively acts as a mass term for the scalar field, and while the geometry is not asymptotically flat, the perturbations admit wave-like asymptotics at large radius. This special feature is lost at the next order ${\cal O}(B^4)$. 
To formalize the expansion, we can introduce a small dimensionless parameter $\epsilon$, scaling $a \rightarrow \epsilon^2 a $ and $B^2 \rightarrow\epsilon B^2$, and then truncate the expansion at $\mathcal{O}(\epsilon^3)$, which retains terms of $\mathcal{O}(a, B^2, aB^2)$. Terms of $\mathcal{O}(B^4)$, while parametrically allowed under this scaling regime, are small inside the Melvin radius for the range of magnetic field considered in this section and will be excluded to ensure Kerr-Newman-like large radius asymptotics as described in Section~\ref{sec:WeaklyMagnetizedEW}.

Under the described scaling regime, at $\mathcal{O}(aB^2)$ the metric components of the Ernst-Wild line element~\ref{e:MagnetizedKNMetric} take the form
\begin{equation}
\label{e:perturbativeEW}
\begin{split}
    g_{tt} &= -\frac{\Delta_r}{ r^2}\left(1+\frac{1}{2}B^2 r^2 \sin ^2\theta \right),\\
    g_{rr} &= \frac{r^2}{ \Delta_r}\left(1+ \frac{1}{2}B^2 r^2 \sin ^2\theta \right),\\
    g_{\theta \theta } &= r^2\left (1 +\frac{1}{2} B^2 r^2 \sin ^2\theta \right ) ,\\
    g_{\phi \phi} &= r^2 \sin ^2\theta \left (1-\frac{1}{2} B^2 r^2 \sin ^2\theta \right ) ,\\
    g_{t \phi} &= -\frac{2a M \sin ^2\theta }{r} \left(1 +2B^2 r^2-\frac{1}{2}B^2 r^2 \sin ^2\theta\right),
\end{split}
\end{equation} 
where $\Delta_r \equiv r(r-2M)$. We consider the Klein-Gordon equation for a massless scalar field, which we denote  $\Phi(t,r,\theta,\phi)$,
\begin{align}
	\label{e:KG_scalar}
    \frac{1}{\sqrt{-g}} \nabla_\mu (\sqrt{-g}g^{\mu \nu} \nabla_\nu \Phi) =0,
\end{align}
and assume a separable ansatz of the following form
\begin{align}
    \label{e:scalarAnsatz}
    \Phi(t, r, \theta, \phi) = e^{- i \omega t} e^{i m \phi} \Phi_r(r) \Phi_\theta(\theta).
\end{align}
The equation governing the radial function $\Phi_r(r)$ is 
\begin{align}
    \label{e:ur}
    \frac{d^2 \Phi_r}{dr_*^2} +(\omega^2 - V)\, \Phi_r = 0, \quad \frac{d}{dr_*} = \frac{\Delta_r}{r^2} \frac{d}{dr},
\end{align}
where $V$ is the effective potential, which can be written as 
\begin{align}
    \label{e:Vr}
    V &= \frac{4  a m M r \omega  \left(2 B^2 r^2+1\right)}{r^4}\nnl
    &- \frac{(2 M-r) \left(B^2 m^2 r^3+\ell (\ell +1) r+2 M\right)}{r^4},
\end{align}
which depends on the frequency of the perturbation. Setting $a=0$ in Eq.~\ref{e:Vr} yields the non-rotating, weakly-magnetized Ernst effective potential, which we will denote $V_E$ and the form can be seen in Fig.~\ref{fig:VrstarErnst} plotted as a function of the tortoise coordinate $r_*$ for $BM = 0.05$. 
\subsection{WKB approximation for quasinormal modes}
Quasinormal modes arise as solutions to a non-Hermitian boundary value problem that describes a linearized perturbation of the black hole geometry. Interpreted as describing the relaxation to equilibrium, they characterize the black hole in terms of its intrinsic properties.  Specifically, the boundary conditions require that modes are purely ingoing at the event horizon ($r_* \rightarrow - \infty$) and purely outgoing as limit $r \rightarrow \infty$ ($r_* \rightarrow + \infty$),
\begin{equation}
    \Phi_r \sim e^{\pm i \omega r_*}, \quad r_* \rightarrow \pm \infty.
\end{equation}
In practice, the above boundary condition at $r\rightarrow \infty$ only applies for asymptotically flat geometries, but as noted above this will also be appropriate for our perturbative treatment of EW.

\begin{figure}
    \centering
    \includegraphics[width=0.9\linewidth]{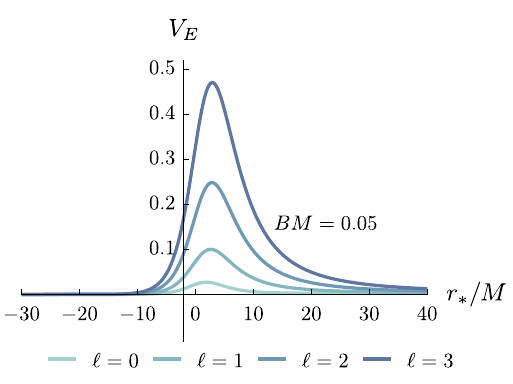}
    \caption{Effective potential $V_E$ of the (weakly-magnetized) Ernst black hole as a function of the tortoise coordinate $r_*$ for $\ell=(0,1,2,3)$. }
    \label{fig:VrstarErnst}
\end{figure}

In this section, we will compute quasinormal modes using the WKB approach, which is generally well-suited to high frequencies, and will be validated against results obtained using Leaver's method in \cite{Taylor:2024duw}. 
Provided that the effective potential $V$ takes the form of a potential barrier with a single peak, the WKB method can be applied. The potential must be constant at $r_* \rightarrow \pm \infty$, although not necessarily approach the same value at both ends and must rise to a maximum at $r_\star = r_0$. This requirement is visually realized in Fig.~\ref{fig:VrstarErnst} which corresponds to the non-rotating limit of the weakly-magnetized Ernst-Wild black hole.
For more detailed discussion on how the WKB procedure is carried out, see, for instance, \cite{Ferrari:1984zz,Iyer:1986np,Iyer:1986nq,Seidel:1989bp,Konoplya:2003ii,Konoplya:2019hlu}. 
The sixth-order WKB method is given by 
\begin{align}
    \label{e:WKB6thOrder}
    \frac{i(\omega^2 - V_0)}{\sqrt{-2 V_0''}}-\sum_{i=2}^6 \Lambda_i = n+ \frac{1}{2}, \quad n \in \mathbb{Z},
\end{align}
where we note that $n= 0, 1, 2, \dots$, for $\text{Re}(\omega) >0$, and $n=-1, -2, -3, \dots$ for $\text{Re}(\omega)<0$.
Here the prime represent derivatives with respect to the tortoise coordinate $r_*$ and $V_0^{(n)}$ denotes the n$^{\text{th}}$ order derivative of the effective potential. The subscript $0$ indicates that the derivatives are to be evaluated at $r_*$ for which the potential obtains its maximum value. The correction terms $\Lambda_2$ and $\Lambda_3$ are given explicitly in \cite{Iyer:1986np} and $ \Lambda_4, \, \Lambda_5$ and $\Lambda_6$ can be found in \cite{Konoplya:2003ii}. They can also be found in a publicly available \texttt{Mathematica} package presented in \cite{Konoplya:2019hlu} for ease of computation. In principle, one needs to determine the position $r_0$ of the maximum of the potential $V(r_0)$, evaluate the various derivatives, and then solve for the QNMs for any value of $n$. The $n=0$ mode is the fundamental mode of oscillation, which turns out to be the least damped mode, and the $n>0$ modes are higher overtones which become more damped with larger imaginary parts as $n$ increases. 

A slight complication arises because the potential, defined in Eq.~\ref{e:Vr}, contains factors of the quasinormal mode frequency $\omega$ within. This obstacle can be overcome by replacing all occurrences of the spin $a$ by $z/\omega$, where $z=a \omega$. Then, all quantities are simply expanded in powers of $z$, which enables the writing of the maximum radius $r_0$ as a function of $a$ and $\omega$ so that the derivatives of the potential can be evaluated to determine the QNMs. The maximum value of the potential proves difficult to determine without first obtaining $r_0$ as a series expansion of this form.
To proceed, we expand the potential $V$ and the quantity $r_0$ as
\begin{align}
    V=V_{E} + V_1(a \omega),
\end{align}
and
\begin{equation}
    r_0 = r_E + r_1(a \omega),
\end{equation}
where $r_E$ is the value of $r$ such that $V_E(r_E)$ is a maximum, corresponding to the weakly-magnetized Ernst potential.  To find $r_0$ we require that to order $(a \omega)$, the following condition must hold
\begin{align}
    V'(r_0) =0,
\end{align}
where all quantities are expanded through order $(a \omega)$. If we demand that the coefficients of all powers of $(a \omega)$ vanish, then we can explicitly calculate the quantities $r_E$ and $r_1$ perturbatively, as explicit functions of $a$ and $\omega$, and $r_0$ is known through order $(a \omega)$. Once $r_0 $ is determined in this way, the master formula~\ref{e:WKB6thOrder} for the quasinormal modes can be computed as a Taylor expansion in powers of $(a \omega)$.  This yields an equation of the form
\begin{align}
    \omega^2 - f(a, n, \omega, \ell, m,  B) =0,
\end{align}
which can then be solved numerically for $\omega$ given $a,\, n, \, \ell, \, m,$ and $B$. For $a=0$, the dependence of the potential on $\omega$ vanishes and so the function $f$ depends solely on $n,\, \ell, \, m,$ and $B$, and determining the Ernst root is simple. For non-zero values of $a$, the function $f$ will be a polynomial in $\omega$. As long as we begin with the Ernst root for $a=0$, we simply follow this root as a starting point for $\omega$ and solve the equation for increasing values of $a$.

\begin{figure}[t]
    \centering
    \includegraphics[width=\linewidth]{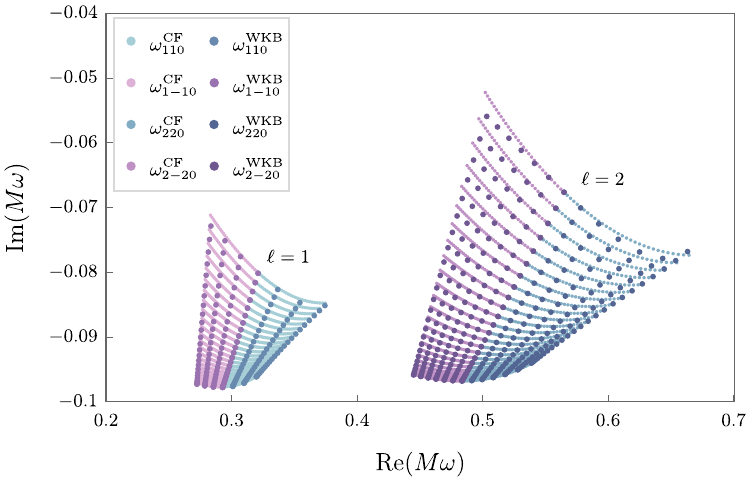}
    \caption{Quasinormal mode frequencies  of the slowly-rotating weakly-magnetized Ernst-Wild black hole are presented for a range of spin $0\leq a \leq 0.3$ and magnetic field $0 \leq B \leq 0.25$. Starting from the bottom, as the spin $a$ increases the lines move in opposite directions horizontally while as $B$ increases the lines move vertically upward. The sixth-order WKB results computed in this work are compared to the Leaver's continued fraction (CF) method results from \cite{Taylor:2024duw}.}
    \label{fig:WKbvsCFResults}
\end{figure}

\begin{figure}[t]
    \centering
    \includegraphics[width=\linewidth]{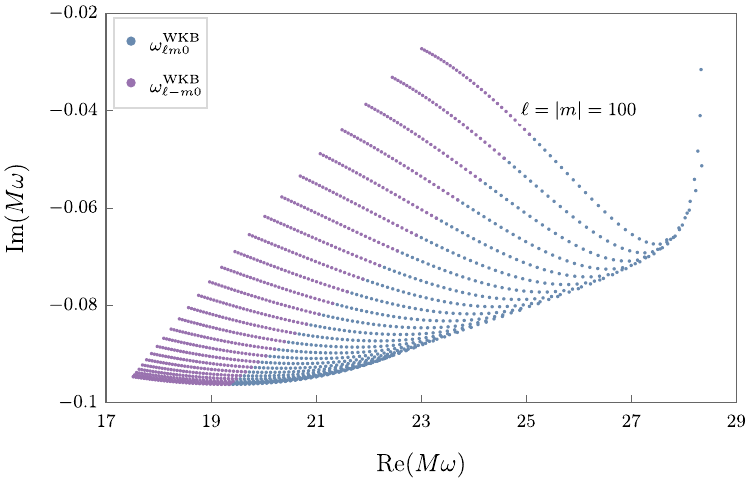}
    \caption{Quasinormal mode frequencies  of the slowly-rotating weakly-magnetized Ernst-Wild black hole are presented using the 6th-order WKB approximation 
    for a range of spin $0\leq a \leq 0.3$ and magnetic field $0 \leq B \leq 0.25$ for $\ell=|m| =100$. }
    \label{fig:largelWKB}
\end{figure}

The results for $\ell = | m | = (1,2)$ are plotted in Fig.~\ref{fig:WKbvsCFResults} and the result match the results obtained using the continued fraction method. Starting from the bottom of the figure, as the spin $a$ increases, the values of the QNMs move in opposite directions horizontally, while as $B$ increases, the modes become more damped and move vertically upward. Increasing the value of the magnetic field $B$ leads to a reduction of the imaginary part of the QNM frequency in all cases. 
This echoes the tendency of the pure Melvin spacetime to simulate confining box-like boundary conditions where normal modes are present as opposed to quasinormal modes.  The points corresponding to Leaver's method exhibit a finer equal spaced mesh $\delta a = \delta B = 0.01$, while the points computed using the sixth-order WKB approximation use $\delta a = 0.1$ and $\delta B = 0.01$. This difference was chosen for practical reasons and enables easier identification between the two methods when the results are displayed together. 
For low spin and magnetic field, the two methods yield nearly identical results but begin to deviate as the spin and magnetic field increases. The points that are at the intersection of the blue and purple points correspond to the Ernst value where the spin $a=0$ and these numerical values do not suffer from the numerical deviations that are clearly present near the edges of the curves when $a\neq 0$ in both directions. This should not come as a surprise, since here the WKB method can be used more reliably since the potential does not depend on the frequency $\omega$. We also plot the results for $\ell = |m| =100$ in Fig.~\ref{fig:largelWKB} to illustrate the behaviour at large $\ell$. The chosen colour scheme will become more readily apparent in the next section when we consider how the QNMs are geometrically related to the light ring. 

\subsection{Light rings in the equatorial plane}
An interesting relation between QNM frequencies in the eikonal limit and unstable null geodesics in the equatorial plane has been studied extensively in the literature; see, for example, the following \cite{Goebel1972ApJ...172L..95G,Ferrari:1984zz,Mashhoon:1985cya,Cardoso:2008bp,Hod:2009td,Dolan:2010wr,Yang:2012he,Dias_2022,Skvortsova:2024atk,Pani_2011}. WKB methods provide an accurate approximation of QNM frequencies in the eikonal (or geometric-optics) limit where one uses the WKB formula without correction terms $\Lambda_i$, namely
\begin{equation}
  \frac{ \omega^2 - V_{0}}{\sqrt{2V''_{0}}} = i(n+1/2).
\end{equation}
Inspection of the above formula leads to the conclusion that in the large $\ell$--limit 
\cite{Ferrari:1984zz,Yang:2012he}, 
\begin{align}
    \label{e:eikonalQNM}
    \omega^\gamma_{\ell m n} \approx \Omega_\gamma (\ell+1/2) - i (n+1/2) |\lambda_\gamma|,
\end{align}
where $\Omega_\gamma$ is the orbital angular velocity of the null circular geodesic, $\lambda_\gamma$ is the principal Lyapunov exponent characterizing the instability of the orbit, $n$ is the overtone number, and $\ell$ is the orbital angular momentum quantum number of the perturbation. We also note that the azimuthal quantum number must be $\ell = |m|$, for the eikonal limit to hold (within the Kerr and Kerr-Newman spacetime).
The Lyapunov exponent, for circular geodesics is given by
\begin{align}
    \label{e:lambdaGamma}
    \lambda_\gamma = \sqrt{\frac{V_r''}{2 \dot t^2}}.
\end{align}
We define a critical exponent as in \cite{Pretorius:2007jn},
\begin{align}
    \gamma \equiv \frac{\Omega_\gamma}{2 \pi \lambda_\gamma} = \frac{T_\lambda}{T_\Omega},
\end{align}
where $T_\Omega \equiv 2\pi/\Omega_\gamma$ is a typical orbital timescale and $T_\lambda$ is an instability timescale as $T_\lambda \equiv 1/\lambda_\gamma$ as in \cite{Cardoso:2008bp}. The critical exponent $\gamma$ is a dimensionless exponent that characterizes the instability of the orbit. 

In this section we denote the quantities associated with the unstable circular light ring of the weakly-magnetized slowly-rotating EW black hole and the Kerr black hole (denoted with subscripts $\gM$ and and $\gK$ respectively). To denote prograde orbits we use the convention that quantities defined with a superscript $+$ are prograde and those defined with a superscript $-$ are retrograde. 

To write down the geodesics for the line element~\ref{e:perturbativeEW} we restrict attention to orbits in the equatorial plane $\theta=\pi/2,\, \dot \theta =0$. Using Eq.~\ref{e:rdotSquared} and the conditions for the existence of circular geodesics, $V_r = V_r' = 0$, lead to the following equations
\begin{equation}
\label{e:rdotsquaredWMSR}
\begin{split}
        V_r =  r_{\gamma_{\text{M}}}^2 E^2(1&-B^2 r_{\gamma_{\text{M}}}^2) + \frac{ (2M -r_{\gamma_{\text{M}}})}{r_{\gamma_{\text{M}}}}L^2 \\
        &-\frac{4 M a \left(1+B^2 r_{\gamma_{\text{M}}}^2\right)}{r_{\gamma_{\text{M}}}}E L = 0,
\end{split}
\end{equation}
and it's derivative,
\begin{equation}
\label{e:VrprimeWMSR}
\begin{split}
        V_r' = 2 r_{\gamma_{\text{M}}}^2& E^2(2 B^2 r_{\gamma_{\text{M}}}^2-1) + \frac{2M}{r_{\gamma_{\text{M}}}}L^2\\
        &+\frac{4 a M \left(B^2 r_{\gamma_{\text{M}}}^2-1\right)}{r_{\gamma_{\text{M}}}}E L =0.
\end{split}
\end{equation}
\begin{figure}[t]
    \centering
    \includegraphics[width=\linewidth]{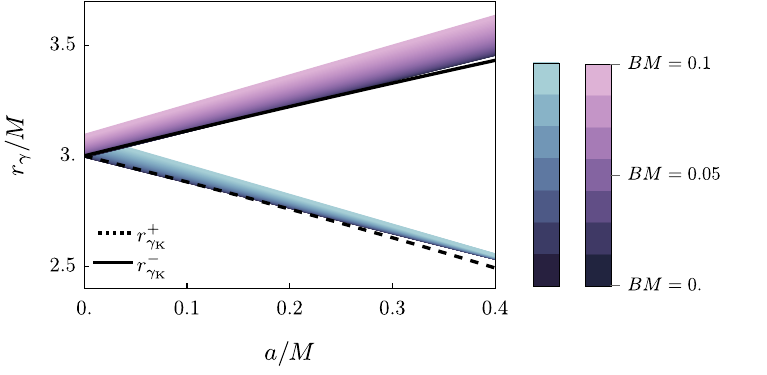}
    \caption{Prograde and retrograde light ring radius as a function of the dimensionless spin parameter $a/M$ for both the slowly-rotating weakly-magnetized Ernst-Wild black hole and the exact Kerr result. }
    \label{fig:rgammaKerrEW}
\end{figure}
Following \cite{Chandrasekhar_theoryofblackholes1983,Cardoso:2008bp,Dolan:2010wr}, the above equations can be simplified by the introduction of a conserved impact parameter $b^\pm_{\gamma} \equiv L/E$, which for weakly-magnetized Ernst-Wild can be written as 
\begin{align}
    \label{e:impactParameter}
    b^\pm _{\gamma_{\text{M}}} =\left(a\pm \frac{(r^\pm_{\gM})^{3/2}}{\sqrt{M}}\right) \left(1-B^2 (r^\pm_{\gM})^2\right),
\end{align}
where $|b^-_{\gM}-a|=-(b^-_{\gM}-a)$ for the retrograde orbit and $|b^+_{\gM}-a|=+(b^+_{\gM}-a)$ for the prograde orbit. We note that $b^{\pm}_{\gM}$ in Eq.~\ref{e:impactParameter} reduces to the Kerr result below on setting $B=0$, 
\begin{align}
    \label{e:impactParameterKerr}
    b^\pm_{\gamma_{\text{K}}} =\left(a\pm\frac{(r^\pm_{\gK})^{3/2}}{\sqrt{M}}\right).
\end{align}
Substituting Eq.~\ref{e:impactParameter} into Eq.~\ref{e:rdotsquaredWMSR} yields the following equation for the radius of circular null geodesics:
\begin{equation}
\begin{split}
3 r_{\gamma_{\text{M}}}^2 -\frac{r_{\gamma_{\text{M}}}^3}{M} &\pm \frac{2 a}{\sqrt{M}}r_{\gamma_{\text{M}}}^{3/2} \pm 8 a B^2 \sqrt{M} r_{\gamma_{\text{M}}}^{5/2} \\
&+B^2 \left(\frac{2 r_{\gamma_{\text{M}}}^5}{M}-5 r_{\gamma_{\text{M}}}^4 \mp \frac{4 a r_{\gamma_{\text{M}}}^{7/2}}{\sqrt{M}}\right) = 0 .
\end{split}
\end{equation}
Solving the above equation within the perturbative expansion yields the radius of null (unstable) circular geodesics and is given by
\begin{align}
    \label{e:rGammaEW}
    r_{\gamma_{\text{M}}}^\pm = 3M \mp \frac{2a}{\sqrt{3}}+ 9 B^2 M^3 \mp 11\sqrt{3} a B^2 M^2.
\end{align}
\begin{figure}[t]
    \centering
    \includegraphics[width=\linewidth]{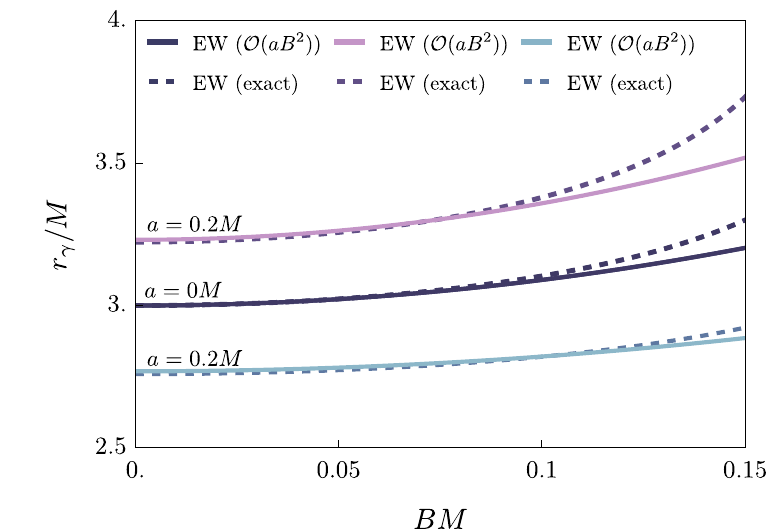}
    \caption{Linear unstable prograde and retrograde light ring orbits compared to the exact results for the spin values $|a/M|=(0, 0.2)$. }
    \label{fig:EW_WMSR_comparisonProRetro}
\end{figure}
\begin{figure*}
    \centering
    \includegraphics[width=\linewidth]{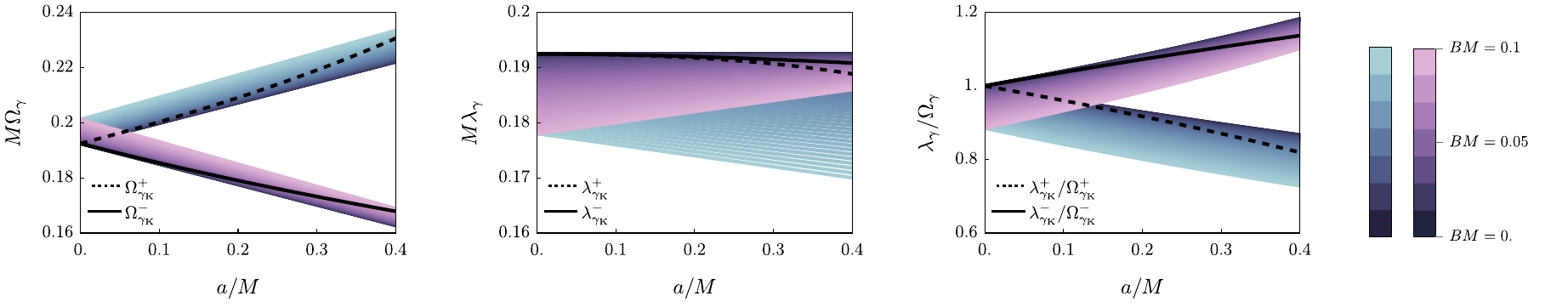}
    \caption{Unstable null circular geodesics associated with the Ernst-Wild black hole in the perturbative regime of $\mathcal{O}(a, B^2)$ and the exact Kerr result with the same mass and spin, shown as a function of dimensionless spin $a/M$. }
    \label{fig:eikonalPlotRange}
\end{figure*}
We compare this result with the Kerr result in Fig.~\ref{fig:rgammaKerrEW}, where the black lines specifically describe the boundary of the exact Kerr photon shell \cite{Johnson:2019ljv} given by 
\begin{align}
    \label{e:rGammaKerrexact}
    r_{\gamma_{\text{K}}}^\pm = 2M \left [ 1+ \cos \left (\frac{2}{3} \arccos\left (\mp \frac{a}{M} \right ) \right ) \right ].
\end{align}
This relation coincides with Eq.~\ref{e:rGammaEW} up to $\mathcal{O}(a)$ when $B=0$, but we actually find
excellent agreement when $B$ is small over a rather wide range of the spin parameter $a$. 
\begin{table*}[]
\centering
\caption{Comparison of the fundamental ($n=0$) quasinormal modes using  $6^{\text{th}}$--order WKB ($\omega_{\ell m n}^{\text{WKB}} $) and using the eikonal limit formula ($\omega^{\gamma}_{\ell m n }$) with $l=m=100$.}
\begin{tabular}{c@{\hspace{1em}}c@{\hspace{1em}}c@{\hspace{1em}}c@{\hspace{1em}}c@{\hspace{1em}}c@{\hspace{1em}}c}
\toprule
\multirow{2}{*}{$a/M$} & \multicolumn{2}{c}{$BM=0.05$} & \multicolumn{2}{c}{$BM=0.1$} & \multicolumn{2}{c}{$BM=0.15$} \\
\cmidrule(lr){2-3} \cmidrule(lr){4-5} \cmidrule(lr){6-7}
  & $\omega_{\ell m n}^{\text{WKB}} $ & $\omega^{\gamma}_{\ell m n }$ & $\omega_{\ell m n}^{\text{WKB}} $ & $\omega^{\gamma}_{\ell m n }$ & $\omega_{\ell m n}^{\text{WKB}} $ & $\omega^{\gamma}_{\ell m n }$ \\[0.5ex]
\midrule
0 & $19.5571 - 0.0944i$ & $19.5588 - 0.0944i$ & $20.2096 - 0.0890i$ & $20.2116 - 0.0890i$ & $21.3153 - 0.0796i$ & $21.2995 - 0.0800i$ \\
0.1 & $20.3791 - 0.0943i$ & $20.3535 - 0.0942i$ & $21.0839 - 0.0897i$ & $21.1570 - 0.0880i$ & $22.2661 - 0.0819i$ & $22.4962 - 0.0777i$ \\
0.2 & $21.3630 - 0.0932i$ & $21.1482 - 0.0939i$ & $22.1353 - 0.0894i$ & $22.1025 - 0.0870i$ & $23.4206 - 0.0832i$ & $23.6929 - 0.0755i$ \\
0.3 & $22.5828 - 0.0898i$ & $21.9429 - 0.0937i$ & $23.4477 - 0.0865i$ & $23.0479 - 0.0860i$ & $24.8782 - 0.0810i$ & $24.8896 - 0.0732i$ \\
\hline
\bottomrule
\end{tabular}
\label{tab:qnm_lm_100}
\end{table*}
A comparison to the exact value (from Fig.~\ref{fig:ProgradeRetrograde2}) is also plotted in Fig.~\ref{fig:EW_WMSR_comparisonProRetro} for small values of the spin parameter, $|a/M|=(0, 0.2)$, where it is understood that blue denotes prograde and purple corresponds to retrograde. 

An important quantity for the analysis of the null geodesics is the angular velocity at the geodesic, 
\begin{equation}
    \label{e:angularVelocityGamma}
    \Omega^\pm_\gamma 
    = \frac{\dot \phi}{\dot t}.
\end{equation}
When the result is evaluated at the light ring $r=r_{\gM}$, we can rewrite Eq.~\ref{e:tdotphidot} in the weakly-magnetized Ernst-Wild spacetime as
\begin{equation}
\begin{split}
    \dot t &= -\frac{r_{\gM } \left(2- B^2 r_{\gM }^2\right)}{4 M-2 r_{\gM }} E + \frac{a M \left(3 B^2 r_{\gM}^2+2\right)}{r_{\gM }^2 \left(2 M-r_{\gM }\right)}L, \\
    \dot \phi &=   \frac{a M \left(3 B^2 r_{\gM }^2+2\right)}{r_{\gM }^2 \left(2 M-r_{\gM }\right)}E - \left(\frac{B^2}{2}+\frac{1}{r_{\gM}^2}\right)L .
\end{split}
\end{equation}
\noindent We now use the metric computed perturbatively up to order $a$ and $B^2$, as in Eq.~\ref{e:perturbativeEW} to compute the orbital frequency and the dimensionless instability for the slowly-rotating, weakly-magnetized Ernst-Wild black hole. 
Substituting Eq.~\ref{e:rGammaEW} into Eq.~\ref{e:angularVelocityGamma} yields the following expansion for the orbital angular frequency defined with respect to coordinate time $t$:
\begin{equation}
    \label{e:OmegaGammaEW}
    \Omega^\pm_{\gM} = \frac{1}{3 \sqrt{3}M} \pm\frac{2a}{27M^2}+\frac{\sqrt{3}B^2M }{2}\pm \frac{2a B^2 M}{3 }.
\end{equation}
Calculation of Eq.~\ref{e:lambdaGamma} also yields the Lyapunov exponent of the light ring geodesic
\begin{equation}
    \label{e:lambdaGammaEW}
    \lambda^\pm_{\gM} = \frac{1}{3\sqrt{3}M}- \frac{5 B^2M}{2 \sqrt{3}}\mp 2 a B^2M.
\end{equation}
The corresponding results for the exact Kerr spacetime are 
\be
 \Omega^\pm_{\gK} = \frac{1}{b^\pm_{\gK}} = \left(a\pm\frac{(r^\pm_{\gK})^{3/2}}{\sqrt{M}}\right)^{-1}
 \ee
 using Eq.~\ref{e:impactParameterKerr} and 
\begin{align}
    \lambda^\pm_{\gK} = \frac{1}{b^\pm_{\gK}}\left (1- \frac{2a}{b^\pm_{\gK}} \right ) \left ( 1- \frac{a}{b^\pm_{\gK}}\right )^{-1/2},
\end{align}
which match the form in \cite{Cardoso:2004nk,Hod:2009td,Dolan:2010wr}. 
The Kerr results for $\Omega^\pm_{\gK},\, \lambda^\pm_{\gK}$ and $\Omega^\pm_{\gK}/\lambda^\pm_{\gK}$ are displayed in black as marked in Fig.~\ref{fig:eikonalPlotRange} for comparison. In the middle figure, we see the decay rate of the prograde mode asymptotically approaching zero, in agreement with the literature \cite{Cardoso:2008bp,Dolan:2010wr}. 

We note that the equality $\Omega^\pm_{\gamma_{\text{K}}}=1/|b^\pm_{\gamma_{\text{K}}}|$ between the angular velocity and the impact parameter evaluated at the light ring, well established in the exact Kerr spacetime \cite{Chandrasekhar_theoryofblackholes1983,Cardoso:2004nk}, is not satisfied for the Ernst-Wild geometry in the perturbative regime. However, one should not necessarily expect this to hold when considering perturbative limits of the geometry. Indeed, in the limit of slow rotation the equality ceases to hold for arbitrary $a$ in the Kerr spacetime, and for this reason it is not expected to hold exactly within the perturbative regime considered here. We note for completeness that in the Schwarzschild limit we recover $r_\gamma^\pm = 3 M$, $b_\gamma^\pm = 3\sqrt{3}M$, and $\Omega_\pm = \lambda_\gamma^\pm = 1/(3\sqrt{3}M)$. 

Finally, we can quantify the goodness of fit in the relation (\ref{e:eikonalQNM}) between the eikonal QNM and the light ring parameters using Eqs.~\ref{e:OmegaGammaEW} and~\ref{e:lambdaGammaEW}. To do so, we can compare the values obtained to those computed using the $6^{\text{th}}$ order WKB approximation from the previous section. The results for $\ell = m =100$ are shown in Table~\ref{tab:qnm_lm_100} for the fundamental mode ($n=0$). We omit the results for the retrograde orbit because the numerical values obey a similar trend. Defining the deviation parameter,
\begin{align}
    \delta \omega_{\ell m n} = \frac{|\omega^{\text{WKB}}_{\ell m n} - \omega^\gamma_{\ell m n } |}{\omega^{\text{WKB}}_{\ell m n}} \times 100 \%.
\end{align}
we find that the real part of $\delta \omega_{\ell m n}$ remains below $8\%$ for all values of spin and magnetic field listed. The imaginary part of $\delta \omega_{\ell m n}$ remains below $9\%$ for $BM=0.05$ and $BM=0.1$ however, for $BM=0.15$, the percentage error increases to $13\%$ for $a/M=0.2$ and to $20\%$ for $a/M =0.3$. 

\section{Concluding remarks}
In this work we have examined null geodesics in the Ernst-Wild geometry, as a model for light trajectories in the near-horizon region of astrophysical black holes with a magnetosphere. In particular, we determined the equatorial light rings for three motivated choices of the seed Kerr-Newman charge, 
including $q=-2a M B + {\cal O}(B^2)$, as is required to render the geometry electrically neutral. The main differing result here compared to the analysis of \cite{Wang_2021} with $q=0$ is the disappearance of prograde unstable light rings at values of the spin much lower than $a/M=1$. 

For magnetic fields of astrophysical interest, namely with $B \ll B_M \sim 1/M$, the scale separation in the geometry allows the near-horizon region to be isolated from the non-asymptotically flat Melvin-like asymptotics at large radius. This motivated us to consider the detailed structure of null trajectories using backward ray-tracing, and to determine the form of the observable shadow and its dependence on $B$. We observe a non-trivial competition between the distortion associated with rotation and the magnetic field. This competition in sphericity of the shadow is shown in Fig.~\ref{fig:shadowDeform2}. 

To explore perturbations of the light ring orbits, we made use of a perturbative expansion in both $B$ and $a$ to render the perturbation equations separable. This allowed for a generalization of the geometric relation between high-frequency quasinormal modes, determined using the WKB method in the eikonal regime, and the orbital frequency and Lyapunov exponents associated with unstable light rings. 

Although observation of the true light ring structure of astrophysical supermassive black holes is beyond the current sensitivity of the Event Horizon Telescope, it may become feasible with next-generation observatories \cite{Lupsasca:2024xhq}. We were motivated to explore the interplay of magnetization and rotation in the observable signatures such as the geometric form of the shadow, and it would be interesting to extend this analysis to consider other environmental effects. In addition, active supermassive black holes are likely to have force-free rather than matter-free magnetospheres. However, realistic GRMHD solutions of this kind are currently only available in numerical form. 

\begin{acknowledgments}
The authors are grateful to Michel Lefebvre and Prateksh Dhivakar for numerous helpful discussions and for reviewing the manuscript. We also thank Luis Lehner and Suvendu Giri for valuable insight and engaging discussions during the Testing Gravity 2025 conference in Vancouver, BC. This work was supported in part by NSERC, Canada. 
\end{acknowledgments}

\appendix 

\counterwithin*{equation}{section}
\renewcommand\theequation{\thesection\arabic{equation}}

\section{Ernst-Wild spacetime}\label{appendix:Full_EW}
In this Appendix, we recall the full structure of the Ernst-Wild geometry. Following \cite{Aliev:1989wz}, the EW metric (with Kerr-Newman seed electric charge $q$, magnetic charge $p=0$, rotation parameter $a$, and mass $M$) can be written as
\begin{align}
    ds^2 = \varrho^2\, \Lambda  \biggr [ &- \frac{\Delta}{\Sigma} dt^2 + \left ( \frac{dr^2}{\Delta} + d \theta^2 \right ) \biggr ]\nnl
    &+ \frac{\Sigma \sin^2 \theta}{\varrho^2\, \Lambda} (\Lambda_0 d\phi - \varpi dt )^2,
\end{align}
where 
\begin{align}
\begin{split}
        \Delta &= r^2 - 2Mr +  a^2 + q^2 , \\
        \varrho^2 &= r^2 +  a ^2  \cos^2 \theta, \\
	    \Sigma &= \left(r^2+a^2\right)^2-a^2 \Delta  \sin ^2 \theta.  
\end{split}
\end{align}
The function $\La$ is defined as follows,
\begin{equation}
    \Lambda = 1 + \frac{\lambda_1 B + \lambda_2 B^2 + \lambda_3 B^3 + \lambda_4 B^4}{\varrho^2}, 
\end{equation}
where, 
\begin{align}
        \lambda_1 &= 2 a q r \sin ^2\theta, \\
        \lambda_2 &= \frac{3}{2} q^2 \left(a^2+r^2 \cos ^2\theta \right)+\frac{1}{2} \Sigma  \sin ^2\theta,\\
        \lambda_3 &= \frac{a q^3 \left(\left(a^2+2 r^2\right) \cos ^2\theta +a^2\right)}{2 r}\nnl
        &-\frac{(a \Delta  q) \left(a^2 \left(\cos ^2\theta +1\right)+r^2 \cos ^2\theta  \left(3-\cos ^2\theta \right)\right)}{2 r}\nnl
        &+\frac{a q \left(a^2+r^2\right)^2 \left(\cos ^2\theta +1\right)}{2 r},\\
        \lambda_4&= \frac{1}{4} a^2 M^2 \left(a^2 \left(\cos ^2\theta +1\right)^2+r^2 \left(\cos ^2\theta -3\right)^2 \cos ^2\theta \right)\nnl
        &+\frac{1}{4} a^2 M q^2 r \sin ^2\theta  \left(\cos ^2\theta -5\right) \cos ^2\theta \nnl
        &+\frac{1}{4} a^2 M r \left(a^2+r^2\right) \sin ^6\theta\nnl
        &+\frac{1}{16} q^4 \cos ^2\theta  \left(a^2 \left(\sin ^2\theta +1\right)^2+r^2 \cos ^2\theta \right)\nnl
        &+\frac{1}{8} q^2 \left(a^2+r^2\right) \sin ^2\theta  \cos ^2\theta  \left(a^2 \left(\sin ^2\theta +1\right)+r^2\right)\nnl
        &+\frac{1}{16} \varrho ^2 \left(a^2+r^2\right)^2 \sin ^4\theta.
\end{align}
Similarly, 
\begin{equation}
    \varpi  = \frac{(2 M r - q^2)\,a + \varpi_1 B + \varpi_2 B^2 + \varpi_3 B^3 + \varpi_4 B^4}{\Sigma},
\end{equation}
where, 
\begin{align}
        \varpi_1 &= - 2 q r \left(a^2+r^2\right), \\
        \varpi_2 &= -\frac{3}{2} a q^2 \left(a^2+\Delta  \cos ^2\theta +r^2\right),\\
        \varpi_3 &=\frac{1}{2} q^3 r \left(\left(3 a^2+r^2\right) \cos ^2\theta-2 a^2\right)- 2a^2 M q^3\nnl
        &+\frac{1}{2} q r \left(a^2+r^2\right) \left(\left(3 a^2+r^2\right) \cos ^2\theta -a^2+r^2\right)\nnl
        &+M q \left(-a^4+r^2 \left(3 a^2+r^2\right) \sin ^2\theta +r^4\right)\nnl
        &+4 a^2 M^2 q r,
\end{align}
and
\begin{align}
        \varpi_4 &= \frac{1}{2} a^3 M^3 r \left(\cos ^4\theta +3\right)\nnl
        &-\frac{1}{8} a q^4 \left(a^2 \left(\cos ^4\theta +1\right)+r^2 \left(\sin ^2\theta +2\right) \cos ^2\theta \right)\nnl
        &-\frac{1}{16} a q^6 \cos ^4\theta -\frac{1}{4} a^3 M^2 q^2 \left(\cos ^4\theta +3\right)\nnl
        &+\frac{1}{16} a q^2 r^2 \left(a^2+r^2\right) \left(3 \cos ^4\theta -6 \cos ^2\theta +1\right)\nnl
        &-\frac{1}{16} a^3 q^2 \left(a^2+r^2\right) \left(\cos ^4\theta +1\right)\nnl
        &-\frac{1}{4} a^5 M^2 \left(\cos ^4\theta +1\right)+\frac{1}{2} a^3 M^2 r^2 \left(3 \sin ^2\theta -2 \cos ^4\theta \right)\nnl
        &+\frac{1}{4} a M^2 r^4 \left(\cos ^4\theta -6 \cos ^2\theta +3\right)\nnl
        &+\frac{1}{4} a M q^2 r \left(2 r^2 \left(3-\cos ^2\theta \right) \cos ^2\theta\right)\nnl
        &-\frac{1}{4} a^3 M q^2 r \left(-2 \cos ^4\theta -3 \cos ^2\theta +1\right)\nnl                
        &+\frac{1}{8} a M r^3 \left(a^2+r^2\right) \left(-\cos ^4\theta+6 \cos ^2\theta +3\right)\nnl
        &-\frac{1}{8} a^3 M r \left(a^2+r^2\right)\left(-3 \cos ^4\theta -6 \cos ^2\theta +1\right)\nnl
        &+\frac{1}{8} a M q^4 r \cos ^4\theta.
\end{align}
The electromagnetic potential is given by
\begin{equation}
    A_\mu = \Phi_0\, dt + \Phi_3 (\Lambda_0\, d \phi - \varpi dt ), 
\end{equation}
where
\begin{equation}
    \Phi_0 = \frac{\Phi_0^0 + \Phi_0^1\, B + \Phi_0^2\, B^2 + \Phi_0^3\, B^3}{4 \Sigma},
\end{equation}
and
\begin{align}
    \Phi_0^0 &= -4 q r \left(a^2+r^2\right)\\
    \Phi_0^1 &= -6 a q^2 \left(a^2+\Delta  \cos ^2\theta +r^2\right)\\
    \Phi_0^2 &= -3 a^2 q (r+2M) +3q r^3 (r^2+4Mr + \Delta \cos^2 \theta) \nnl
            &- 6 a^2 q^3(r+2M) +18 a^2 q M r^2 +24 M^2 q a^2 r\nnl
            &+9 q a^2 r \Delta \cos^2 \theta,\\
        \Phi_0^3 &= -\frac{a}{2} \Biggr \{4 a^4 M^2+2 a^4 M r+a^4 q^2-24 a^2 M^3 r\nnl
    &+12 a^2 M^2 q^2 -24 a^2 M^2 r^2+4 a^2 M q^2 r+2 a^2 q^4\nnl
    &- 6 \Delta  r \cos ^2\theta  \left(2 M \left(a^2+r^2\right)-q^2 r\right)+4 a^2 M r^3\nnl
    &-12 M^2 r^4 -6 M r^5-q^2 r^4 \nnl
    &+ a^2 \Delta  \cos ^4\theta   \left(4 M^2-6 M r+q^2\right)\nnl
\
        &+ \Delta  \cos ^4\theta  \left(2 M r^3+q^4-3 q^2 r^2\right)
\
    \Biggr \}.
\end{align}
Finally, we also have 
\begin{equation}
        \Phi_3 = \chi = \frac{\chi_0 + \chi_1 B + \chi_2 B^2 + \chi_3 B^3}{\Lambda  \varrho ^2},
\end{equation}
where
\begin{align}
        \chi_0 &= a q r \sin ^2\theta \\
        \chi_1 &=\frac{1}{2} \left(3 q^2 \left(a^2+r^2 \cos ^2\theta \right)+\Sigma  \sin ^2\theta \right)\\
        \chi_2 &= \frac{3}{2} a M q \left(a^2 \left(1-\cos ^2\theta \right)+r^2 \cos ^2\theta  \left(3-\cos ^2\theta \right)\right)\nnl
        &+\frac{3}{4} a q r \left(a^2+r^2\right) \sin ^4\theta -\frac{3}{4} a q^3 r \sin ^2\theta  \cos ^2\theta \\
        \chi_3 &= \frac{1}{2} a^2 M^2 \left(a^2 \left(\cos ^2\theta +1\right)^2+r^2 \cos ^2\theta  \left(3-\cos ^2\theta \right)^2\right)\nnl
        &-\frac{1}{2} a^2 M q^2 r \sin ^2\theta  \left(5-\cos ^2\theta \right) \cos ^2\theta \nnl
        &+\frac{1}{2} a^2 M r \left(a^2+r^2\right) \sin ^6\theta\nnl
        &+\frac{1}{8} \varrho ^2 \left(a^2+r^2\right)^2 \sin ^4\theta \nnl
        &+\frac{1}{8} q^4 \cos ^2\theta  \left(a^2 \left(2-\cos ^2\theta \right)^2+r^2 \cos ^2\theta \right)\nnl
        &+\frac{1}{4} q^2 \left(a^2+r^2\right) \sin ^2\theta  \cos ^2\theta  \left(a^2 \sin ^2\theta +a^2+r^2\right).
\end{align}

\section{Apparent horizons}
\label{app:apparenthorizon}
In this Appendix we present the calculation determining the apparent horizon geometry of both the Kerr and Ernst black hole, leading to Fig.~\ref{fig:ernstKerrAH} of the main text. 
\begin{figure*}
    \centering
    \includegraphics[width=\linewidth]{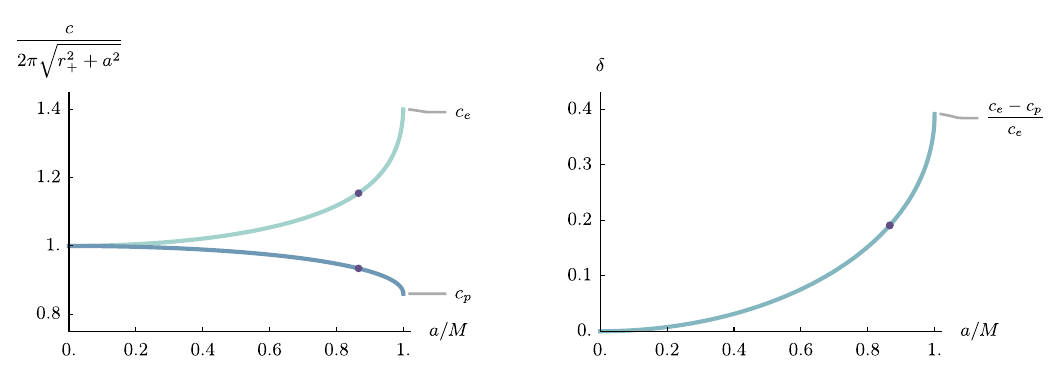}
    \caption{Proper equatorial circumference $c_e$ and proper polar circumference $c_p$ for the Kerr black hole. The purple dots on the curves at $a/M = \sqrt{3}/2$, represent the point above which the surface can not be globally embedded in Euclidean 3--space. The value $\delta = (c_e-c_p)/c_e$ specifies how increasingly oblate the surface is becoming.}
    \label{fig:kerrGrid_AH}
\end{figure*}
\subsection{Kerr black hole}
The induced metric on the horizon for a Kerr black hole is given by \cite{Smarr:1973zz}
\begin{align}
    \label{e:inducedKerrmetric}
    ds_{(2)}^2 = (r_+^2 + a^2 \cos^2 \theta)\, d\theta^2 + \frac{(r_+^2 + a^2 )^2\, \sin^2 \theta}{r_+^2 + a^2 \cos^2 \theta}\, d\phi^2,\!
\end{align}
where $r_+$ is the event horizon of the Kerr black hole and is defined in the usual way,
\begin{equation}
    r_+ = M^2 -\sqrt{M^2 -a^2}.
\end{equation}
To obtain a rough measure of the surface asphericity, it is useful to compare the equatorial circumference $c_e$ ($\theta = \pi/2$) and the polar circumference $c_p$ ($\phi =0$). Calculation yields
\begin{align}
    c_e = \int_{0}^{2 \pi} \frac{r_+^2 +a ^2}{r_+}\, d\phi = \frac{2 \pi\, (r_+^2 +a ^2)}{r_+},
\end{align}
and
\begin{align}
    c_p = \int_{0}^{2 \pi} \sqrt{r_+^2 + a^2 \cos^2 \theta}\, d \theta = 4 \sqrt{r_+^2 + a^2}\, E(\beta),
\end{align}
where for simplicity we define the function $\beta$ as
\begin{equation}
    \beta^2 = \frac{a^2}{r_+^2 +a^2}.
\end{equation}
In the above solution for the polar circumference, $E(\beta)$ denotes the complete elliptical integral of the second kind. These circumferences are invariant quantities since the curves are geodesics of the 2--metric.

\begin{figure*}
    \centering
    \includegraphics[width=\linewidth]{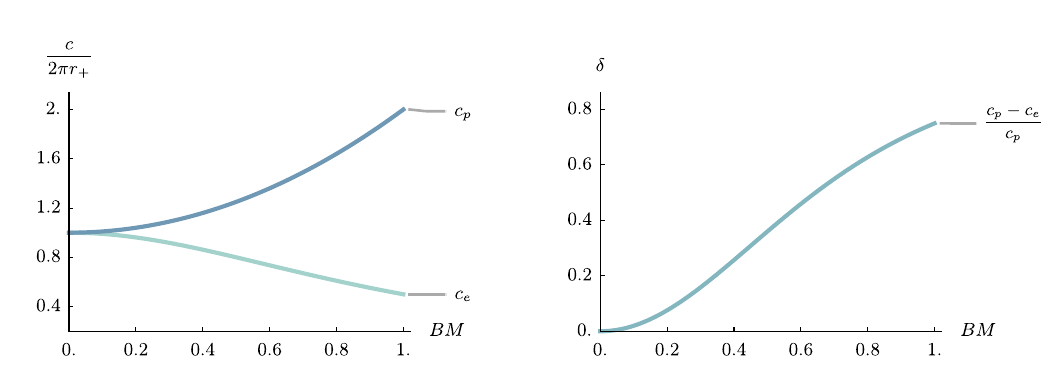}
    \caption{Proper equatorial circumference $c_e$ and proper polar circumference $c_p$ for the Ernst black hole. Another measure of the surface distortion of the Ernst black hole due to the magnetic field. The value $\delta = (c_p-c_e)/c_p$ specifies how increasingly prolate the surface is becoming. }
    \label{fig:ernstGridAH}
\end{figure*}

To emphasize the oblate deviation we introduce the deviation parameter 
\begin{equation}
    \label{e:deltaKerr}
    \delta = \frac{c_e - c_p}{c_e} = 1-\frac{2}{\pi } \sqrt{\frac{r_+^2}{a^2+r_+^2}} E\left(\beta \right).
\end{equation}
The proper equatorial circumference $c_e$ and the proper polar circumference $c_p$ for the Kerr geometry are plotted in Fig.~\ref{fig:kerrGrid_AH} alongside the deviation $\delta$.  Since the metric~\ref{e:inducedKerrmetric} is of the form $ds_{(2)}^2 = F(\theta)^2\, d \theta^2 + G(\theta)^2\, d \phi^2$,
the Gaussian curvature of the intrinsic geometry is given by  $K = -\dfrac{1}{2 FG} \dfrac{d}{d \theta} \left ( \dfrac{1}{FG} \dfrac{d}{d\theta}G^2\right )$.
\noindent Calculation of the above expression for the Kerr black hole yields
\begin{equation}
    K = \frac{\left(r_+^2+a^2\right) \left(r_+^2-3 a^2 \cos ^2\theta \right)}{\left(r_+^2+a^2 \cos ^2\theta \right)^3}.
\end{equation}
This quantity measures the geometry intrinsic to the horizon itself and is independent of the chosen embedding space. To check the topology of the surface $\Sigma$, one simply makes use of the Gauss-Bonnet theorem \cite{dray2014differential}, which states that
\begin{align}
    \label{e:GaussBonnet}
    \int_\Sigma K d A + \int_{\partial \Sigma}\kappa_g\,  ds= 2 \pi  \chi(\Sigma),
\end{align}
where $\chi(\Sigma)$ is the Euler characteristic of the surface $\Sigma$ and $\kappa_g$ is the geodesic curvature of $\partial \Sigma$. The second integral in the above form vanishes because the curves are geodesics, and by definition it must be $\kappa_g \equiv 0$. 
At the event horizon, we determine the following double integral as
\begin{equation}
    \int_{0}^{2 \pi}\int_0^{\pi} K \sqrt{r_+^2 +a^2} \sin \theta \, d \theta \, d \phi  = 4 \pi,
\end{equation}
and comparing with Eq.~\ref{e:GaussBonnet}, we identify that the black hole surface is topologically a 2--sphere with $\chi(\Sigma)=2$.
The surface area of the black hole event horizon is an invariant property of interest and is given by 
\begin{equation}
    A_+ = \int_{0}^{2 \pi} \int _0^\pi  \sqrt{r_+^2 +a^2} \sin \theta \, d \theta\, d \phi = 4 \pi\, (r_+^2+a^2).
\end{equation}
When $a/M=\sqrt{3}/2$, the Gaussian curvature becomes zero at the poles $\theta =0$. Beyond this value, a global embedding in Euclidean 3--space is impossible. 

Finally, in order to visualize the intrinsic geometry isometrically in Euclidean 3--space we follow the standard procedure \cite{Smarr:1973zz} as follows. The mapping $(\theta, \phi) \mapsto (x, y, z)$ is given by,
\begin{align}
    x = \rho(\theta) \cos \phi,\, \quad y = \rho(\theta) \sin \phi, \quad z= z(\theta).
\end{align}
The induced cylindrical metric is obtained by pulling back the Euclidean metric and can be written as 
\begin{align}
    \label{e:cylindricalAnsatzInduced}
    ds^2_{(2)} = \left (\rho '(\theta)^2 +  z '(\theta) ^2 \right )\, d \theta^2 + \rho(\theta)^2\, d \phi^2.
\end{align}
We must match the metric \ref{e:inducedKerrmetric} with the above result which yields
\begin{align}
    \rho(\theta) &= \frac{(r_+^2 + a^2 )\, \sin \theta}{\sqrt{r_+^2 + a^2 \cos^2 \theta}}, \\
    z(\theta) &= \int \sqrt{(r_+^2 + a^2 \cos^2 \theta) - \left ( \frac{d \rho}{d \theta}\right )^2 } \, d \theta.
\end{align}
The above equation for $z(\theta)$ must then be integrated numerically. 
In the Kerr spacetime, in the left image of Fig.~\ref{fig:ernstKerrAH} depicting the Euclidean embedding diagram in the $x$--$z$ plane, we see that for $a=0$, the horizon is a perfect circle of radius $r=2M$. This matches the Schwarzschild horizon. For moderate spin, $0 < a/M < \sqrt{3}/2$: as the spin increases, the shape becomes oblate, meaning that it flattens along the $z$--axis and elongates along the $x$--axis. Embedding in flat space must then stretch along the equatorial direction and shrink along the polar direction - this keeps the integrated surface area of the horizon $A_+$ equal - this is a consequence of how the curved, intrinsic geometry is represented in a flat Euclidean diagram. As Smarr showed in \cite{Smarr:1973zz}, a global embedding of the Kerr horizon is possible only if $a/M\leq \sqrt{3}/2$, as the Gaussian curvature becomes negative above this value.

\subsection{Ernst black hole}
The Ernst line element can be found by setting $a=0$ and $q=0$ into Eq.~\ref{e:MagnetizedKNMetric} of the main text, 
\begin{align}
    ds^2 = &- \Lambda^2 \left (1- \frac{2M}{r} \right ) dt^2 + \Lambda^2 \left (1- \frac{2M}{r} \right )^{-1} dr^2 \nnl
    \label{e:ErnstLineElement}
    &+ r^2 \Lambda^2 d \theta^2 + \frac{r^2 \sin^2 \theta}{\Lambda^2 } d \phi^2,
\end{align}
where $\Lambda$ is simplified to 
\begin{align}
    \Lambda = 1+  \frac{B^2 r^2 \sin^2 \theta}{4}.
\end{align}
The induced 2--metric on the outer horizon for the Ernst black hole is given by
\begin{align}
    ds_{(2)}^2 = r_+^2 \Lambda_+^2 d \theta^2 + \frac{r_+^2 \sin^2 \theta}{\Lambda_+^2 } d \phi^2,
\end{align}
where 
\begin{align}
    \Lambda_+ = 1+  \frac{B^2 r_+^2 \sin^2 \theta}{4}.
\end{align}
Similarly to the Kerr apparent horizon procedure, we must identify the equatorial and polar circumference $c_e$ and $c_p$. Calculations yield
\begin{align}
    c_e = \int_0^{2 \pi} r_+\, \Lambda_+\, d\phi = \frac{8 \pi  r_+}{4+B^2 r_+^2}
\end{align}
and similarly,  
\begin{align}
    c_p = \int_0^{2 \pi} \frac{r_+}{\Lambda_+}\,\sin \theta\, d\phi = \frac{\pi\, r_+ \left(4+B^2 r_+^2\right)  }{2} .
\end{align}
Here to emphasize the prolate deviation we define $\delta$ as
\begin{align}
    \delta = \frac{c_e - c_ p}{c_p} = 1-\frac{1}{\left(B^2+1\right)^2}.
\end{align}
The Gaussian curvature for the Ernst black hole is given by 
\begin{align}
    K = &\frac{\Lambda_+ + 2 B^2 M^2}{r_+^2  \Lambda_+^3} - \frac{12 M^4 B^4 \sin^2 \theta \cos^2 \theta}{r_+^2 \Lambda_+^4}.
\end{align}
Setting the magnetic field strength $B=0$ recovers the result for the Schwarzschild  black hole, that is, $K = 1/r_+^2$, which is simply the curvature of a sphere of radius $r_+$. 
The surface area of the event horizon of the Ernst black hole reduces to that of Schwarzschild and is given by 
\begin{align}
    A_+ = \int_{0}^{2 \pi} \int_0^{\pi} r_+^2 \sin \theta \, d\theta\, d \phi = 4 \pi r_+^2.
\end{align}
In the same way as Kerr, it is simple to show that the surface is topologically a sphere, once again through the use of the Gauss-Bonnet theorem given by Eq.~\ref{e:GaussBonnet}. Calculation identically yields $\chi(\Sigma) =2$ - indicating once again that the event horizon is topologically a 2--sphere. 
As for Kerr, we plot the proper equatorial and polar circumferences in Fig.~\ref{fig:ernstGridAH}.  Here we notice that as $B$ increases, the equatorial circumference decreases since $\Lambda_+$ occurs in the denominator. In contrast, the polar circumference increases as $B$ increases, since $\Lambda_+$ now occurs in the numerator. 
Finally, the Euclidean embedding diagram for the Ernst black hole can be seen in the right-hand side of Fig.~\ref{fig:ernstKerrAH} of the main text following the same steps for Kerr from the cylindrical ansatz given in Eq.~\ref{e:cylindricalAnsatzInduced}. 
\bibliography{refs.bib}
\end{document}